\documentclass[aps,pra, twocolumn,superscriptaddress,floatfix,nobibnotes]{revtex4-1}
\usepackage{amsmath,amsfonts,amssymb,bbold}
\usepackage{graphicx}
\usepackage{epstopdf}
\usepackage[english]{babel}
\usepackage[utf8]{inputenc}

\usepackage{letltxmacro}

\LetLtxMacro{\ORIGselectlanguage}{\selectlanguage}
\makeatletter
\DeclareRobustCommand{\selectlanguage}[1]{%
  \@ifundefined{alias@\string#1}
    {\ORIGselectlanguage{#1}}
    {\begingroup\edef\x{\endgroup
       \noexpand\ORIGselectlanguage{\@nameuse{alias@#1}}}\x}%
}
\newcommand{\definelanguagealias}[2]{%
  \@namedef{alias@#1}{#2}%
}
\makeatother

\definelanguagealias{en}{english}
\definelanguagealias{eng}{english}
\definelanguagealias{English}{english}

\let\originalleft\left
\let\originalright\right
\renewcommand{\left}{\mathopen{}\mathclose\bgroup\originalleft}
\renewcommand{\right}{\aftergroup\egroup\originalright}

\newcommand{\bra}[1]{\ensuremath{\left\langle #1\right|}}
\newcommand{\ket}[1]{\ensuremath{\left|#1\right\rangle}}

\newcommand{\braket}[2]{\left\langle #1,#2\right\rangle}

\providecommand{\abs}[1]{\left\lvert#1\right\rvert}

\renewcommand{\phi}{\varphi}

\begin{document}

\title[Scattering Theory of Efficient Quantum Transport across Finite Networks]{Scattering Theory of Efficient Quantum Transport across Finite Networks}

\author{Mattia Walschaers}
\affiliation{Physikalisches Institut, Albert-Ludwigs-Universit\"at  Freiburg, Hermann-Herder-Str. 3, D-79104 Freiburg, Germany}
\affiliation{Instituut voor Theoretische Fysica, KU Leuven, Celestijnenlaan 200D, B-3001 Leuven, Belgium}
\affiliation{Laboratoire Kastler Brossel, UPMC-Sorbonne Universit\'e, CNRS, ENS-PSL Research University, Coll\`ege de France}
\author{Roberto Mulet}
\affiliation{Department of Theoretical Physics, Physics Faculty, University of Havana, La Habana, CP 10400, Cuba.}
\author{Andreas Buchleitner}
\email{andreas.buchleitner@physik.uni-freiburg.de}
\affiliation{Physikalisches Institut, Albert-Ludwigs-Universit\"at  Freiburg, Hermann-Herder-Str. 3, D-79104 Freiburg, Germany}

\begin{abstract}
We present a scattering theory for the efficient transmission of an excitation across a finite network with designed disorder. 
We show that the presence of randomly positioned networks sites allows to significantly accelerate the excitation transfer processes 
as compared to a dimer structure, if only the disordered Hamiltonians are 
constrained to be centrosymmetric, and to exhibit a dominant doublet in their spectrum. We identify the cause of this 
efficiency enhancement in the constructive interplay between disorder-induced fluctuations of the dominant doublet's splitting 
and the coupling strength between the input and output sites to the scattering channels. We find that the characteristic 
strength
of these fluctuations together with the channel coupling fully control the transfer efficiency.
\end{abstract}

\maketitle

\section{Introduction}

Excitation transport across finite, discrete and disordered networks \cite{weber_transport_2010,levi_quantum_2015,doi:10.1146/annurev-conmatphys-031115-011327}
defines an abstract model for a variety of quantum transport problems, 
with applications to realistic physical scenarios which reach from natural or artificial light harvesting \cite{Spallek} to the physics of cold Rydberg gases \cite{PhysRevLett.108.023005, scholak_spectral_2014,Deng:2016aa}. 
Beyond fundamental aspects of disorder-induced localisation on discrete networks, this setting also defines an interesting incidence of quantum control
in the presence of (and, possibly, through) disorder, which, by the very nature of disordered systems, enforces a statistical approach. Furthermore,
when it comes to the specific context of light harvesting, one faces 
finite networks embedded into hierarchical super-structures \cite{FMO_NatChem, Kramer17}, with hitherto only barely understood interfacing.
For a conceptual  understanding of the necessary structural elements which ensure the functionality of light harvesting units, it is
indispensable to elucidate how the associated, broadly distributed length, energy and time scales are orchestrated, and how the different elementary 
building blocks are interconnected \cite{Ringsmuth2012}.

As a first elementary ingredient  
of a properly equipped toolbox for 
the modular modelling of such hierarchical structures, our present contribution establishes 
a scattering theoretical description of statistical control of point-to-point quantum transport across disordered, finite networks. 
We adopt the 
perspective that an excitation, collected e.g. by the antenna complex of a photosynthetic functional unit, is injected into the network at a 
specific input site, and extracted at an output site from 
where it is channeled towards the reaction centre -- the sub-unit where the incoming photon's energy is used to drive
the ATP cycle \cite{FMO_NatChem}.
We elucidate the 
statistics of the resonance structures in the associated transmission cross sections of the network, together with the concomitant excitation transfer times. 
In particular, we investigate the interplay between network structure and effective coupling to input and output leads, under the specific, 
coarse grained constraints of centrosymmetry and the presence of dominant doublet states in the networks' spectra. 

\section{Model}

We build our model as a network described by a Hilbert space $\mathbb{C}^N$ (given $N$ network sites). To interconnect the network with the 
structures it is embedded in, 
it is attached to scattering channels (or leads) 
which are described by a space ${\cal H}_c \subset \mathcal{L}^2(\mathbb{R}^3)$ (because we generally consider wave propagation in three 
spatial dimensions). We can thus describe the full Hilbert space of the network 
together with the external channels as ${\cal H}_{\rm total} = \mathbb{C}^N \bigoplus_c {\cal H}_c.$
Physically this means that we consider a set of bound states
on the network, which are coupled to a continuum --
a setting familiar from nuclear and atomic physics \cite{fano_effects_1961,feshbach_unified_1958,feshbach_unified_1962,feshbach_unified_1967}.
An incoming excitation is described as 
a wave packet which is coupled into the network via one of the 
leads, and subsequently creates an excitation 
on the network. This excitation will 
subsequently decay, with variable delay, into either one of the 
leads. 

\subsection{Transfer Probability and Dwell Times}

For a rigorous description of this scenario we introduce 
the scattering matrix 
as defined\footnote{Direct application of the formalism of \cite{feshbach_unified_1958,feshbach_unified_1962,feshbach_unified_1967} leads to 
$$S(E)=\mathbb{1}- 2 \pi i \hat{W}^{\dag}\frac{1}{E - H_{\rm eff}}\hat{W}\, ,$$ hence 
$W = \sqrt{\pi}\hat{W}.$ Both
conventions appear in the literature, 
and we 
choose (\ref{eq:Smatrix}) for notational convenience.} \cite{brouwer_quantum_1997, celardo_superradiance_2009,haake_statistics_1992,lewenkopf_stochastic_1991,seba_statistical_1996,stockmann_effective_2002} by: 
\begin{equation}\label{eq:Smatrix}\begin{split}
S(E)&\equiv\mathbb{1}- 2 i W^{\dag}\frac{1}{E - H_{\rm eff}}W,\\
 &\text{with}\quad H_{\rm eff} = H-i WW^{\dag},
\end{split}
\end{equation}
where $H$ is the network's Hamiltonian
represented by an $N \times N$ matrix. The latter encodes  
the relative positions of the network's nodes (e.g., a set of chlorophyll molecules) and their mutual couplings. 
The scattering matrix depends on the energy of the incoming particle, $E$, and 
has a dimension given by
the number $n_c$ of attached channels. Also the operator $W$ is 
described by a matrix, but of dimension $n_c \times N$, 
with entries which determine the coupling between leads and 
network sites. Consequently, 
$H_{\rm eff}$ is a non-Hermitian \cite{rotter_non-hermitian_2009} $N \times N$ matrix, 
while $S(E)$ has
dimension $n_c \times n_c$.

Note that every scattering channel does itself 
support a continuum of modes with 
continuously distributed energies.  
Thus, there may be an additional energy dependence of $W$, representing how different modes of 
the same channel couple to the network. However, 
we here assume that such energy dependence can be 
omitted -- a common approximation
both in light-matter interactions \cite{cohen-tannoudji_atom-photon_1998} 
and in mesoscopic physics \cite{lewenkopf_stochastic_1991}. 
Furthermore, the present formulation (\ref{eq:Smatrix}) of the scattering matrix assumes the Lamb shift \cite{cohen-tannoudji_atom-photon_1998} to 
be negligible. 

We now 
quantify our control target, the {\em excitation transfer efficiency across the network}, in the present scattering theoretical 
context. We 
define two figures of merit: the {\em transfer probability} $p_{c\rightarrow c'}(E)$, and the {\em dwell time} $\tau_{c\rightarrow c'}(E)$ --
from channel $c$ to channel $c'$ --, which 
both 
depend
on the injection energy $E$.
The transfer probability 
reads 
\cite{brouwer_quantum_1997}
\begin{equation}\label{eq:PScatter}
p_{c\rightarrow c'}(E) \equiv \abs{S_{c,c'}(E)}^2\, ,
\end{equation}
and the {\em dwell time} \cite{berkolaiko_moments_2010,brouwer_quantum_1997,kuipers_semiclassical_2008,seba_statistical_1996,smith_lifetime_1960} is 
given by \cite{brouwer_quantum_1997,smith_lifetime_1960}
\begin{equation}\label{eq:DefTauDenkIK}
\tau_{c \rightarrow c'} (E) \equiv {\rm Im} \left\{ {S_{c,c'}(E)}^{-1} \frac{\rm d}{{\rm d}E}S_{c,c'}(E) \right\}\, .
\end{equation}
The latter quantity denotes the phase shift
imprinted on an incoming plane wave 
during the scattering process, and can be interpreted as the time needed for the incoming wave packet to be scattered into the output lead.
Let us add that the {\em resonance lifetimes} \cite{Rotter2013}, determined by the imaginary parts $\Gamma_i/2$ of the 
resonance eigenvalues ${\cal E}_i = E_{i} - i \Gamma_i/2$ of $H_{\rm eff}$, provide another set of time scales which 
characterize 
the scattering process. 
As long as distinct resonances 
do not overlap, i.e., $\Gamma_i,\Gamma_j \ll \abs{E_i-E_j}$ for all $i$ and $j$, dwell times and resonance lifetimes are intimately related. However, for overlapping resonances, such direct association (see e.g.~\cite{lyuboshitz_collision_1977}) breaks down. 


\subsection{Design Principles for Efficient Transfer}\label{sec:desprinc}

In a next step, we rely on earlier results which identified centrosymmetric \cite{zech_centrosymmetry_2014,walschaers_optimally_2013,walschaers_statistical_2015,walschaers_currents,ortega_quantum_2015,PhysRevE.94.042102} 
random networks as more efficient than unconstrained 
random assemblies, and specify 
our scattering model as given by centrosymmetric Hamiltonians of the form
\begin{equation}\label{eq:HCentro}
H=\begin{pmatrix}
E'  &  v_1 \dots  v_n & V\\
v_1 &  & v_n \\
\vdots &  H_{\rm int}& \vdots \\
v_n & & v_1 \\
 V &  v_n \dots  v_1 & E' 
\end{pmatrix},
\end{equation}
with $H_{\rm int}$ a centrosymmetric matrix which represents the hardwiring 
between the 
bulk sites of the network. $H$ commutes with the exchange operator $J$, where $J_{ij} = \delta_{i, N-j+1}$ in the site basis. $E'$ gives the on-site 
energy of the input and of the output site of the network, coupled with strengths $v_i$ to the bulk sites, and with $V$ to each other. Only input and output
site be coupled to input and output channel, respectively, and $\ket{\rm in}$ and $\ket{\rm out}$ be 
states fully localised on these 
respective sites, with $\ket{\rm in} = J \ket{\rm out}$. If we represent the input and output channel states by 
%
$\ket{\Psi_{\rm in}}$ and $\ket{\Psi_{\rm out}}$, respectively, we can 
express the coupling operator $W$ in (1) as 
\begin{equation}
W = \sqrt{\frac{\Gamma}{2}}\ket{{\rm in}}\bra{\Psi_{\rm in}} + \sqrt{\frac{\Gamma'}{2}}\ket{{\rm out}}\bra{\Psi_{\rm out}}\, ,
\end{equation}
and $H_{\rm eff}$ in (\ref{eq:Smatrix}) thus takes the form 
\begin{equation}\label{eq:HeffCentro}
H_{\rm eff}= H - i \frac{\Gamma}{2} \ket{\rm in}\bra{\rm in}- i \frac{\Gamma'}{2} \ket{\rm out}\bra{\rm out}.
\end{equation}
%
%
With the additional choice $\Gamma=\Gamma'$ \cite{walschaers_currents}, $H_{\rm eff}$ becomes a {\em non-Hermitian centrosymmetric matrix}. 

\begin{figure}[t]
  \centering
\includegraphics[width=0.45\textwidth]{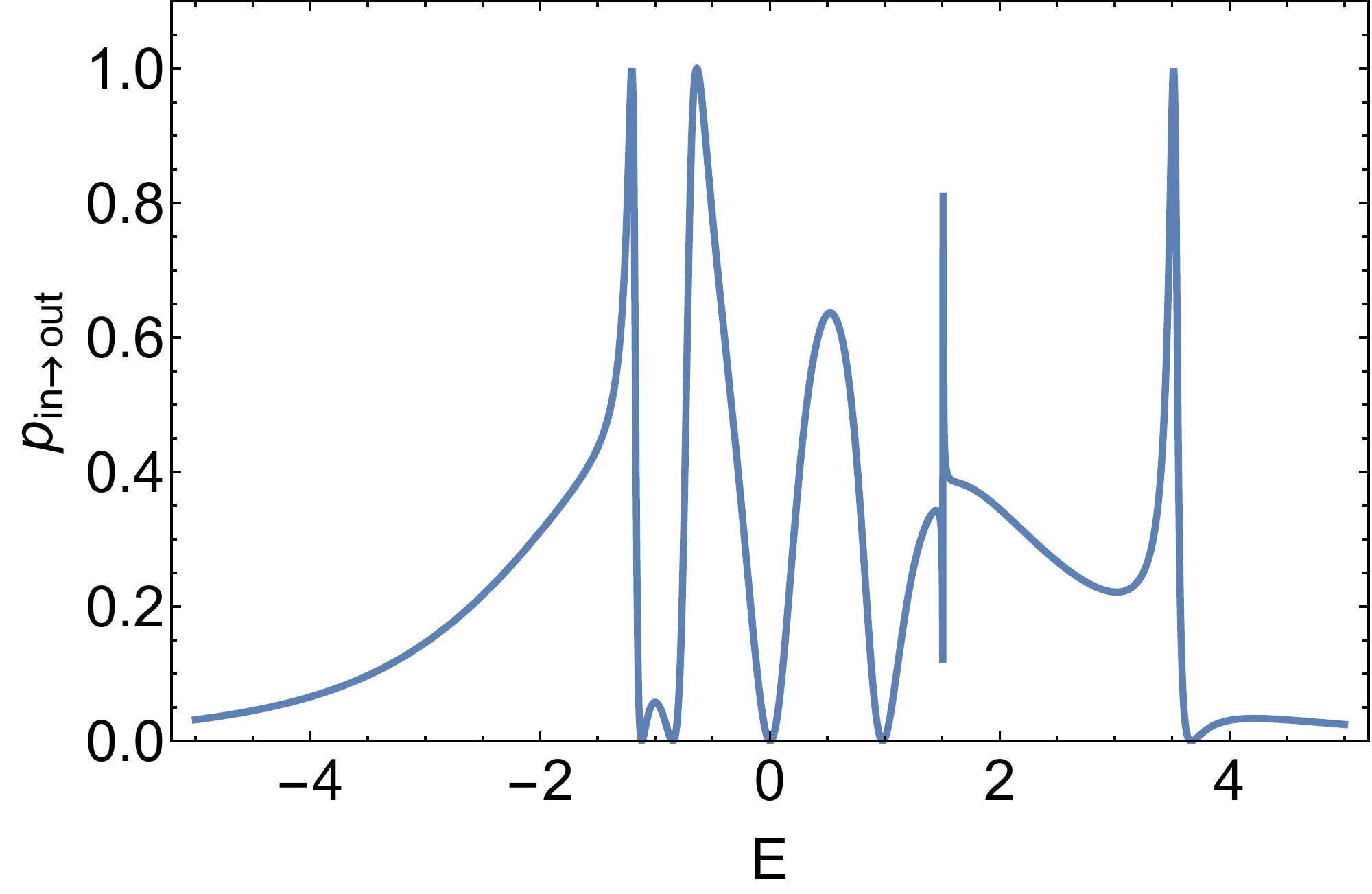}
\caption{Resonance profile, as given by transfer probability (\ref{eq:PScatter}) for a single, randomly chosen network Hamiltonian $H$ (\ref{eq:HCentro}). Couplings between input/output sites and the bulk, $v_i$, follow the same statistics as intermediate site couplings $(H_{\rm int})_{jk}$, and are sampled according to Section \ref{sec:NumericsForScattering}. In (\ref{eq:HeffCentro}) and (\ref{eq:hierstaatdexi},\ref{eq:hierstaatdechi}) we set $\chi=\xi = 1$, $\Gamma = 5$, $E' = 0$, and $V = 1$. The system contains a total of $N=8$ sites. Figured obtained from \cite{Walschaers_Thesis}.}
\label{fig:ResCentro}
\end{figure}

\begin{figure}[t]
  \centering
\includegraphics[width=0.45\textwidth]{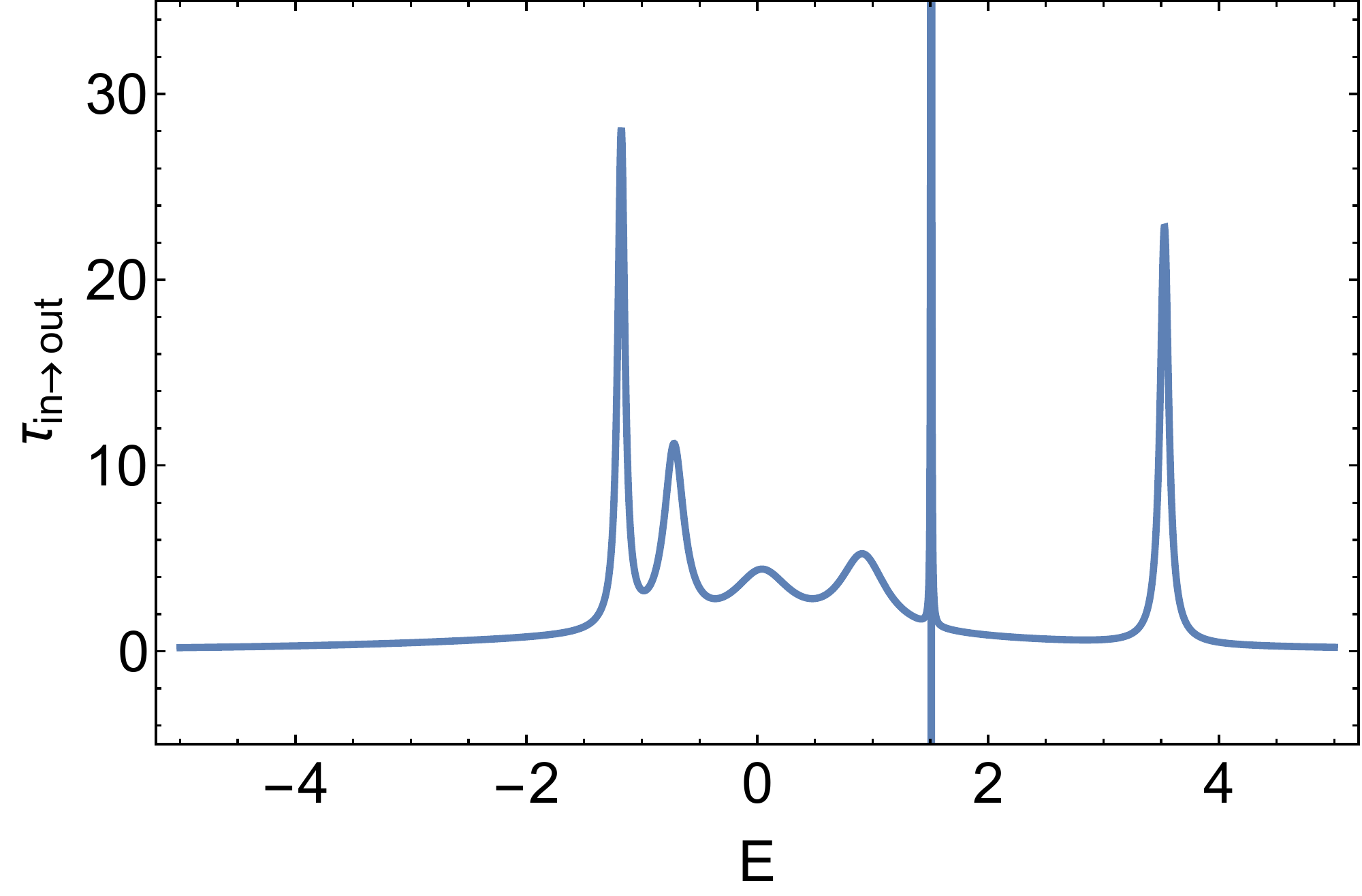}
\caption{Dwell time profile (\ref{eq:DefTauDenkIK}) for the same $H$ as was sampled for Fig.~\ref{fig:ResCentro}. Figure obtained from \cite{Walschaers_Thesis}.}
\label{fig:DwellTimeCentro}
\end{figure}

However, centrosymmetry alone is not yet sufficient to guarantee efficient transport features of the networks, as illustrated in Figs. 1 and 2, where the 
transfer probability $p_{\rm in \rightarrow out}$ and the dwell time $\tau_{\rm in \rightarrow out}$ are plotted for a typical, centrosymmetric random network
of $N=8$ sites.
In Fig.~\ref{fig:ResCentro}, we observe
asymmetric 
(e.g.~at $E\approx 1.5$, due to the interference of overlapping resonances) as well as 
symmetric resonance structures, with variable 
widths and strengths, some of which achieve 
$p_{\rm in \rightarrow out} = 1$. 
Yet, Fig.~\ref{fig:DwellTimeCentro} 
highlights that large resonant transfer probabilities may be associated with -- here undesirable -- very long time scales.
This is consistent with the narrow resonance structures in Fig.~\ref{fig:ResCentro}, which imply long 
resonance lifetimes,\footnote{Also note the negative value of $\tau_{\rm in \rightarrow out}$ at  $E\approx 1.5$, which hints at the fact that 
(\ref{eq:DefTauDenkIK}) actually quantifies a phase shift rather than a genuine time.} as well as with earlier findings \cite{walschaers_optimally_2013,walschaers_statistical_2015}.

We therefore need to introduce another design principle, the {\em dominant doublet} condition -- which requires  
that the network Hamiltonian $H$ exhibits eigenstates close to 
\begin{equation}
\ket{\pm}\equiv \frac{1}{\sqrt{2}}(\ket{\rm in} \pm \ket{\rm out})
\end{equation} 
More formally, given the centrosymmetric matrix (\ref{eq:HeffCentro})
in the symmetry eigenbasis 
(in which $J$ is diagonal),
\begin{equation}\label{eq:MatrixComplexApp}
H_{\rm eff} = \begin{pmatrix} E'+V-i \frac{\Gamma}{2}&\bra{\mathcal{V}^+}& &  \\ \ket{\mathcal{V}^+} & H^+_{sub}& & \\
& & E'-V -i \frac{\Gamma}{2} & \bra{\mathcal{V}^-}\\ & & \ket{\mathcal{V}^-} &H_{sub}^- \end{pmatrix}. \end{equation}
the dominant doublet condition
imposes the existence of 
two eigenvectors $\ket{\eta^{+}}$ and $\ket{\eta^-}$ such that 
\begin{equation}
\label{eq:doubletCond}
\abs{\braket{\eta^{\pm}}{\pm}}^2 = 1 - \epsilon\, ,
\end{equation} 
with $\epsilon >0$ close to zero. 

With the help of lowest order perturbation theory \cite{walschaers_optimally_2013,walschaers_statistical_2015}, (\ref{eq:doubletCond}) allows to derive analytical approximations for our quantities
of interest. To start with, for the eigenvalues associated with 
$\ket{\eta^{+}}$ and $\ket{\eta^-}$ one finds:
\begin{equation}\label{eq:EigPertScatter}
\begin{split}
& {\cal E}^{\pm} \approx E' \pm V -i \frac{\Gamma}{2}+s^{\pm},\\ &\quad {\rm where} \quad s^{\pm}=\sum_i \frac{\abs{\braket{\mathcal{V}^{\pm}}{\psi^{\pm}_i}}^2}{E' \pm V - e^{\pm}_i},
\end{split}
\end{equation}
where $e^{\pm}_i$ and $\ket{\psi^{\pm}_i}$ are the eigenvalues and eigenvectors of $H_{sub}^{\pm}$ in (\ref{eq:MatrixComplexApp}), respectively.
Setting all $v_i = 0$ in (\ref{eq:HCentro}) implies $s^{\pm} = 0$, and 
reproduces the result for a 
network shrunk to a dimer composed only of 
input and output site.

Combining Eqs.~(\ref{eq:doubletCond}) and (\ref{eq:EigPertScatter}) we now determine the 
scattering matrix (\ref{eq:Smatrix}) which maps the input to the output channel as
\begin{align}
S_{{\rm in} \rightarrow {\rm out}}(E) 
&= -i \frac{\Gamma}{2}\left( \frac{1}{E - {\cal E}^{+}} - \frac{1}{E - {\cal E}^{-}} \right) + {\cal O}(\epsilon)\, , \label{eq:ScatteringMatrixAddedSites}
\end{align}
from which, with (\ref{eq:EigPertScatter}), we infer 
the transfer probability:\footnote{We will henceforth omit the index ``${\rm in}\rightarrow {\rm out}$'', for ease of notation.}
\begin{equation}\label{eq:HierIsDeEne}
\begin{split}
p(E) \approx &\frac{\frac{\Gamma^2}{4} (2 V + \Delta s)^2}{\left( (E'-V +s^- -E)^2+\frac{\Gamma^2}{4}\right)} \\
&\qquad\times\frac{1}{\left( (E'+V+s^{+}-E)^2+\frac{\Gamma^2}{4}\right)},
\end{split}
\end{equation}
where 
$\Delta s = s^+ - s^-$. 
Similarly, $\tau(E)$ is derived from (\ref{eq:EigPertScatter}) with the help of 
(\ref{eq:DefTauDenkIK}), 
leading to the somewhat cumbersome expression:
{\small \begin{equation}\begin{split}\label{eq:HierIsDeAndere}
\tau(E)\approx &\Bigg\{\frac{\Big(\Gamma^2+4 {E'}^2+4 E' \left(s^-+s^+-2 E\right)\Big)}{\left(\Gamma^2+4 \left(E'+s^--V-E\right)^2\right)}\\
&\qquad \times\frac{4 \Gamma}{\left(\Gamma^2+4 \left(E'+s^++V-E\right)^2\right)}\Bigg\}\\
&+\Bigg\{\Bigg(\frac{-2 s^- (V+E)+{s^-}^2+{s^+}^2}{\left(\Gamma^2+4 \left(E'+s^--V-E\right)^2\right)}\\
&\qquad+\frac{2 \left(V \left(s^++V\right)-s^+ E+E^2\right)}{\left(\Gamma^2+4 \left(E'+s^--V-E\right)^2\right)}\Bigg)\\
&\qquad \times\frac{8 \Gamma}{\left(\Gamma^2+4 \left(E'+s^++V-E\right)^2\right)}\Bigg\}.
\end{split}
\end{equation}}

\begin{figure}[t]
  \centering
  \includegraphics[width=0.45\textwidth]{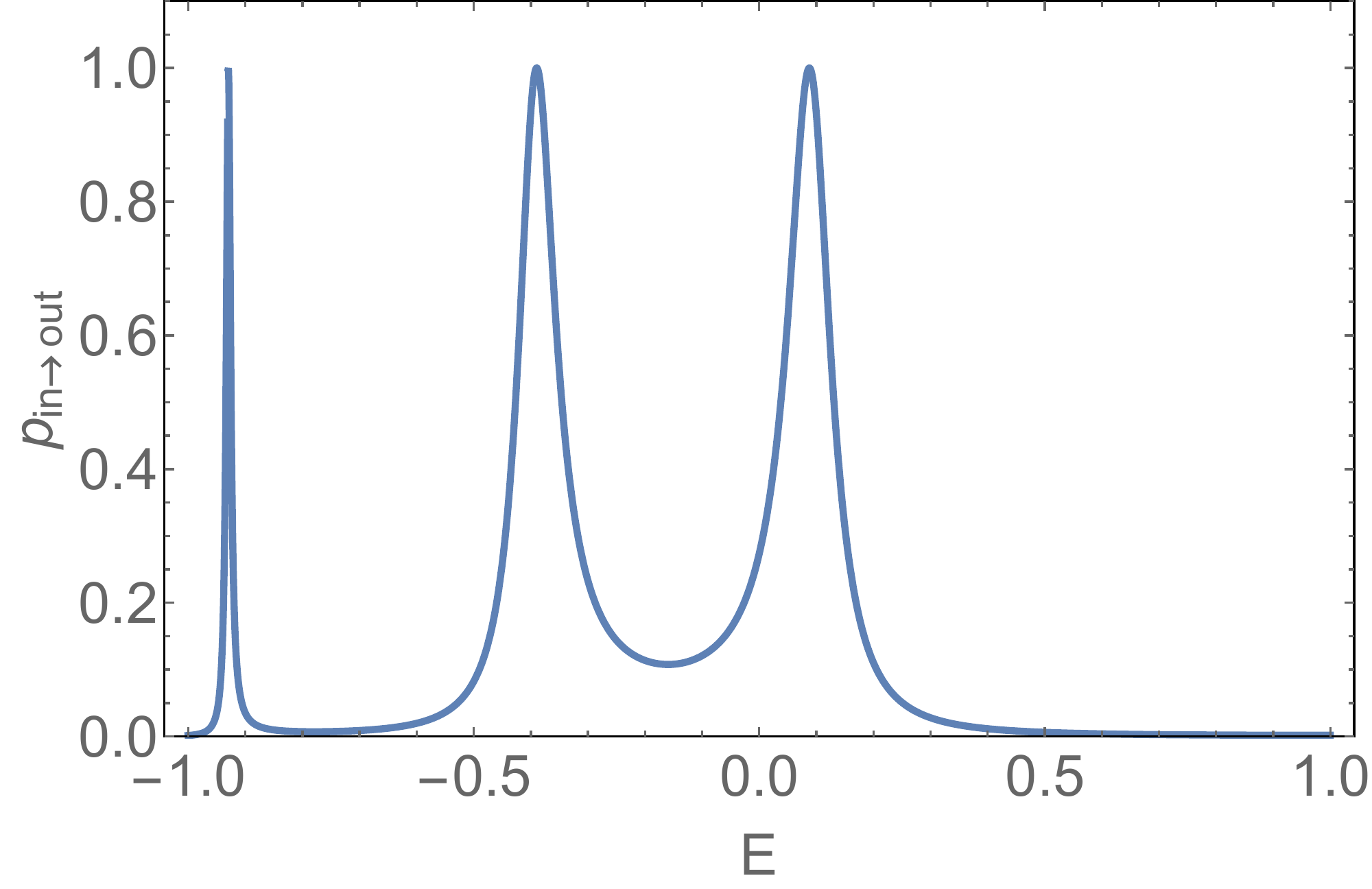}
\includegraphics[width=0.45\textwidth]{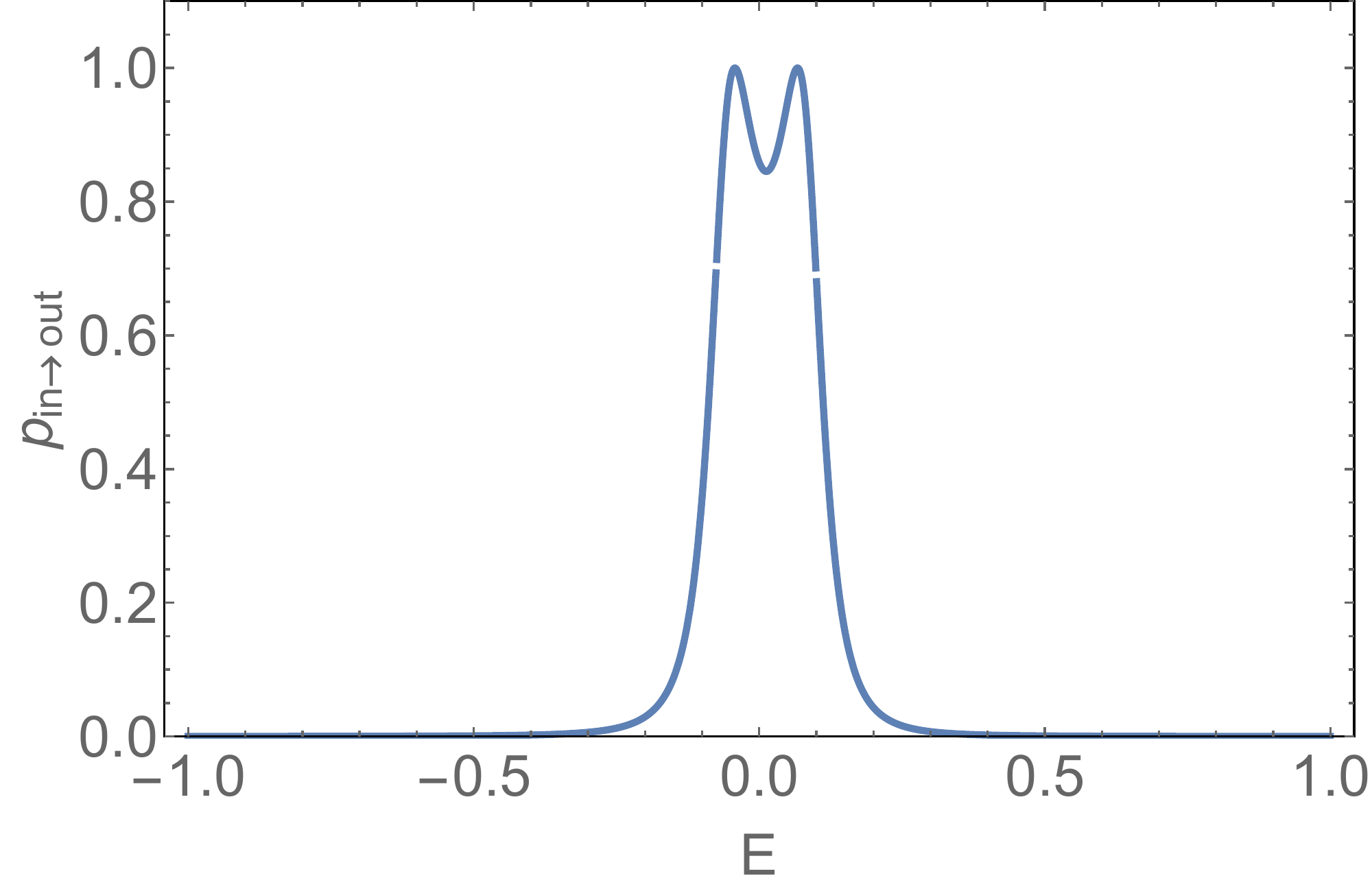}
\includegraphics[width=0.45\textwidth]{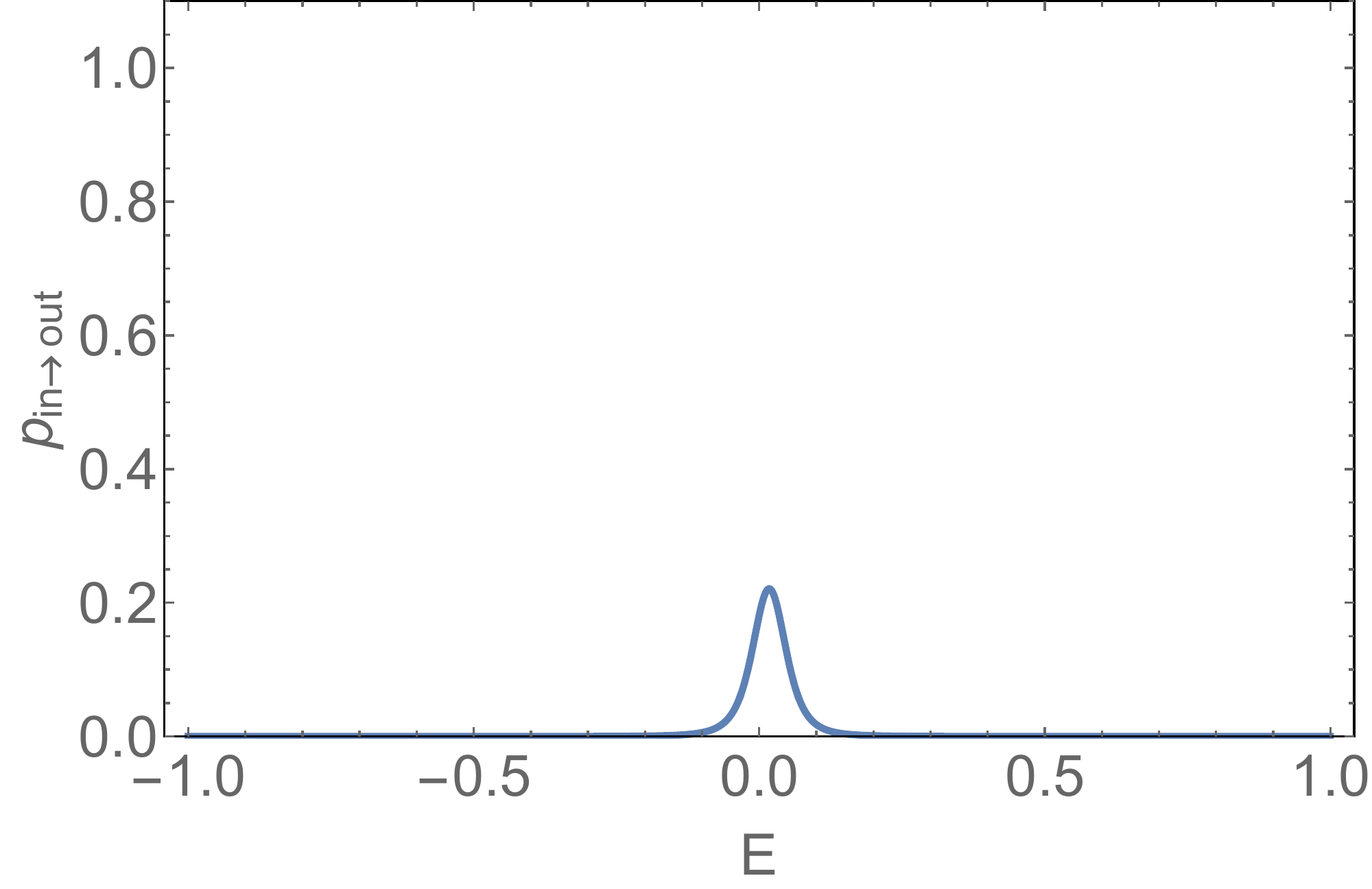}
\caption{Resonance profiles for three randomly chosen network Hamiltonians $H$, (\ref{eq:HCentro}). These three networks of $N=10$ sites are sampled following Section \ref{sec:NumericsForScattering}, where we set $\xi = 10$ (\ref{eq:hierstaatdexi}) and $\chi=1$ (\ref{eq:hierstaatdechi}), and keep $E' = 0$, $V = 0.01$, and $\Gamma = 0.2$ fixed. Shown are realisations with $\abs{2V+ \Delta s} = 0.484257$ (top), $\abs{2V+ \Delta s} = 0.14596$ (middle), and $\abs{2V+ \Delta s}= 0.024839$ (bottom). Note that in the top panel, an additional resonance peak 
induced by the intermediate sites is visible.}
\label{fig:ResCentroDoub}
\end{figure}

Inspection of (\ref{eq:HierIsDeEne}) and (\ref{eq:HierIsDeAndere}) shows that transfer probability and dwell time 
sensitively depend 
on the energy $E$ of the incoming excitation. 
From (\ref{eq:HierIsDeAndere}),
the energies which maximise the transfer probability $p$ are
\begin{align}
&E = E' + \overline{s}\, ,{\rm for}\ \Gamma\geq\abs{2V + \Delta s}\, ,\label{eq:PResNLevel2} \\
&E =  E' + \overline{s} \pm\frac{1}{2} \sqrt{(2 V + \Delta s)^2 -\Gamma^2} \, ,{\rm for}\ \Gamma < \abs{2V + \Delta s}\, ,\label{eq:PResNLevel}
\end{align}
with $\overline{s} \equiv (s^+ + s^-)/2$ 
the average shift of the resonance energy.
The transfer efficiency at these resonant energies follows as 
\begin{equation}\label{eq:approxpP}
\begin{split}
&p(E'+\overline{s}) \approx \frac{\Gamma^2}{V^2}\frac{(2V +\Delta s)^2}{\left(\Gamma^2 + (2V+\Delta s)^2\right)^2},\\
&p\left(E' + \overline{s} \pm \frac{1}{2} \sqrt{(2 V + \Delta s)^2 -\Gamma^2}\right) \approx  1,
\end{split}
\end{equation}
and the associated dwell times are 
\begin{equation}\label{eq:approxT}
\begin{split}
&\tau(E'+\overline{s}) \approx \frac{4 \Gamma}{\Gamma^2+(2V + \Delta s)^2},\\
&\tau\left(E' + \overline{s} \pm \frac{1}{2} \sqrt{(2 V + \Delta s)^2 -\Gamma^2}\right) \approx  \frac{2}{\Gamma}.
\end{split}
\end{equation}
Eq.~(\ref{eq:approxpP}) implies that, whenever $\Gamma < \abs{2V + \Delta s}$, there are two energies given by Eq.~(\ref{eq:PResNLevel}), at which 
the incoming wave packet is transmitted deterministically.
Consequently, the profile of $p$ as given by (\ref{eq:HierIsDeEne}) exhibits two well-separated resonances, as shown in the top 
panel of Fig.~\ref{fig:ResCentroDoub}. 
For $\Gamma = \abs{2V + \Delta s}$, the two resonances 
merge, and the maximal transfer probability is achieved at $E=E' + \overline{s}$ (Fig.~\ref{fig:ResCentroDoub}, middle panel).
The transfer probability starts to decrease
as the ratio between $\Gamma$ and $\abs{2V + \Delta s}$ increases beyond unity, as illustrated 
in the bottom panel of Fig.~\ref{fig:ResCentroDoub}.

The on-resonance dwell times $\tau$ in (\ref{eq:approxT}) are governed by the parameter $\Gamma$. Indeed, it directly follows from (\ref{eq:approxT}) that, on resonance, $2/\Gamma < \tau < 4/\Gamma$. Therefore, to obtain fast transport at large transfer probabilities, we must make $\Gamma$ as large as possible, under the constraint that $\abs{2V + \Delta s} > \Gamma$. 

In the above perturbative approach, the impact of the
bulk sites of the network is absorbed 
in the 
shifts $\Delta s$ and $\overline{s}$ of 
the dominant doublet.
Apart from these shifts, the system then effectively behaves as a two-level system, comprised only of the input and output sites. 
Of course, there are additional resonances, at other energies than those of the doublet, 
as clearly visible in the top panel of Fig.~\ref{fig:ResCentroDoub}. These 
{\em must}
have very long life times, since 
the imaginary parts of the associated complex eigenvalues of $H_{\rm eff}$ determine their widths, and only
emerge at 
higher orders of the perturbative expansion (\ref{eq:EigPertScatter}).
While not accounted for in 
the analytical theory as 
presented here, such resonances are fully included in our numerical simulations and have no significant impact on 
the results presented in Sec.~\ref{sec:NumericsForScattering} below.

\section{Statistical Analysis}\label{sec:Results}

Given our above analysis of the resonance structure 
with associated transmission characteristics of single realisations of
disordered, finite networks, we can now address the statistical properties of the transfer probability which results from sampling over 
a distribution of such networks, under the above constraints of centrosymmetry and a dominant doublet. For this purpose, we 
assume that the direct input-output coupling $V$, as well as the channel coupling strength $\Gamma$ have identical values for all
realisations of the network structure, and introduce the scaled system parameters  
 \begin{equation}
 \label{eq:newVars}
\tilde{\Gamma}\equiv \frac{\Gamma}{2V} \quad \, ,\, \quad \Delta \tilde{s}\equiv \frac{\Delta s}{2V}\, .
\end{equation}
Under this assumption, fluctuations of transfer probability and dwell time have their origin in the fluctuations of the relative level shift
$\Delta s$ of the dominant doublet. That correction's distribution function is given by \cite{Lopez1981,walschaers_optimally_2013,walschaers_statistical_2015}
\begin{equation}
\label{eq:Cauchy}
P(\Delta \tilde{s}) = \frac{1}{\pi}\frac{\tilde{\sigma}}{\tilde{\sigma}^2 +(\Delta \tilde{s} - \tilde{s}_0)^2}\, ,
\end{equation}
with $\tilde{\sigma}$ and $\tilde{s}_0$ determined by the mean level spacing and the average coupling strength 
between the dominant doublet and the bulk states. 

We saw above (recall top panel of Fig.~3) that efficient excitation transfer across the network can be achieved for $\Gamma$ as large as possible, 
yet under the condition of well-separated resonances associated with the dominant doublet states. In terms of our scaled
system parameters, this latter condition reads  
\begin{equation}
\label{eq:constraint}
\tilde{\Gamma} < \abs{1+\Delta \tilde{s}}\, ,
\end{equation}
and we can evaluate the probability to fulfill this condition, given (\ref{eq:Cauchy}), as follows:
\begin{align}
{\rm Prob}&(\tilde{\Gamma} < \abs{1 + \Delta \tilde{s}}) \\&= {\rm Prob} (\Delta \tilde{s} > \tilde{\Gamma} - 1 ) + {\rm Prob} (\Delta \tilde{s} < - 1- \tilde{\Gamma}  ) \nonumber \\
&= 1 - \int ^{\tilde{\Gamma} - 1}_{ - \tilde{\Gamma} - 1} {\rm d}\Delta \tilde{s}\,P(\Delta \tilde{s})\\
&=1 - \frac{1}{\pi}\arctan \left( \frac{\tilde{\Gamma}-1-\tilde{s}_0}{\tilde{\sigma}}\right)\label{eq:densEffReal}\\
&\nonumber\qquad\quad- \frac{1}{\pi}\arctan \left( \frac{\tilde{\Gamma}+1+\tilde{s}_0}{\tilde{\sigma}}\right).
\end{align}

\begin{figure}[t]
  \centering
\includegraphics[width=0.49\textwidth]{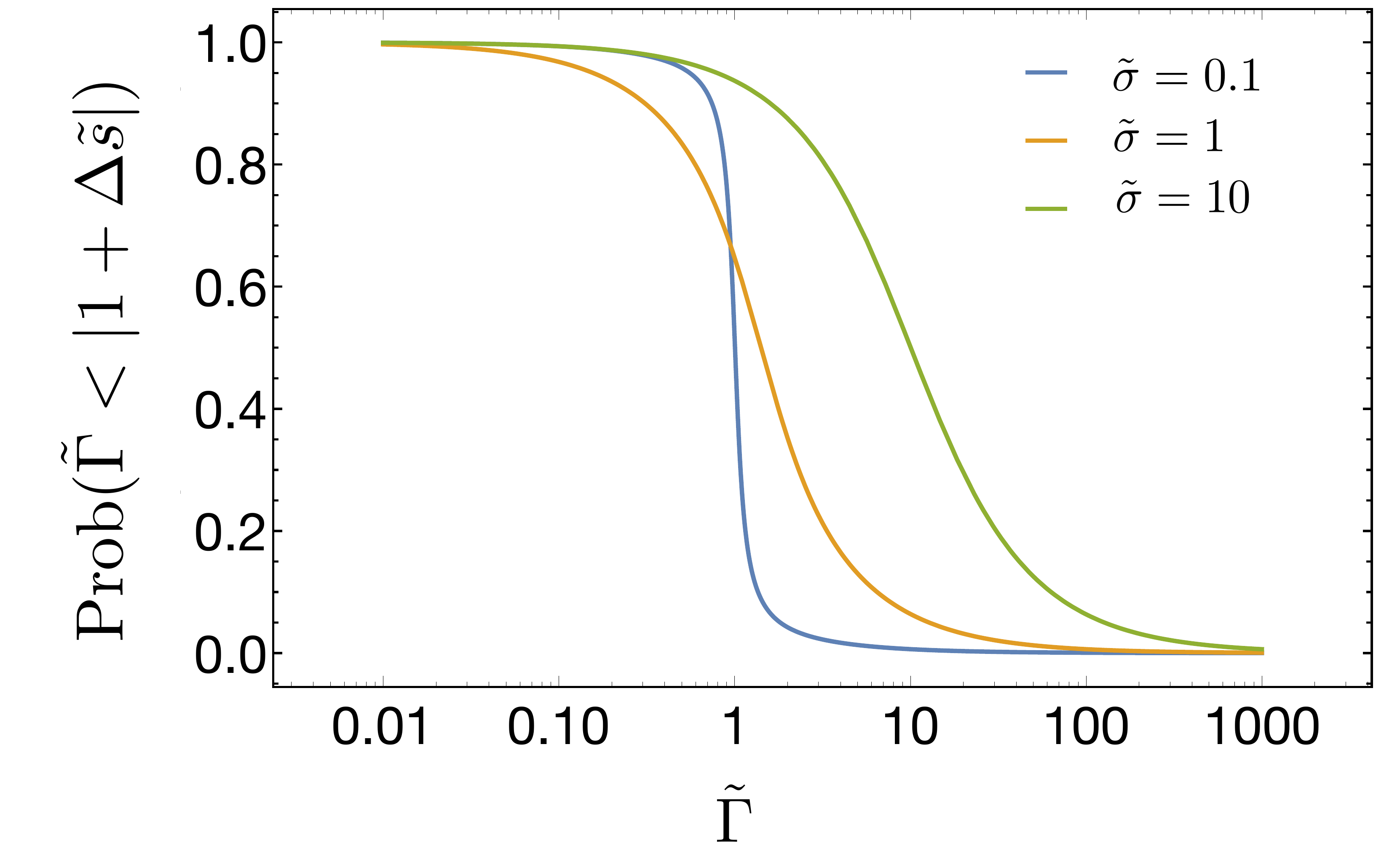}
\caption{Density of efficient realisations (\ref{eq:densEffReal}), as a function of rescaled coupling to the channels $\tilde{\Gamma}=\Gamma/2V$. Shown are three different values of the width of the distribution of energy shifts, relative to the input-output coupling, $\tilde{\sigma}=\sigma/2V$. The parameter $\tilde{s}_0=s_0/2V$ is set to zero, which is consistent with the dominant doublet condition (\ref{eq:doubletCond}, \ref{eq:domdubstat}). Figure obtained from \cite{Walschaers_Thesis}.}
\label{fig:densEffReal}
\end{figure}

The density (\ref{eq:densEffReal}) is plotted in Fig.~\ref{fig:densEffReal}, for different widths $\tilde{\sigma}$ of the distribution of $\Delta \tilde{s}$, 
while we set $s_0=0$ (what is justified since 
$s_0\ll 1$ in general \cite{walschaers_statistical_2015}, if the dominant doublet condition (\ref{eq:doubletCond}) is fulfilled).
There is a clear transition 
from deterministically separated dominant doublet resonances for $\tilde{\Gamma}\ll 1$, to overlapping resonances for $\tilde{\Gamma}\gg 1$. For $\tilde{\sigma}\ll 1$, this change occurs drastically at $\tilde{\Gamma}\approx 1$. However, the transition is smoothened, and shifted towards larger 
values of $\tilde{\Gamma}$, with increasing values of $\tilde{\sigma}$. Furthermore, overlap of the doublet resonances becomes likely for 
$\tilde{\Gamma} \sim 1+\tilde{\sigma}$, and large $\tilde{\sigma}$ is therefore desirable,
since enhanced values of $\tilde{\Gamma}$ which still comply with (\ref{eq:constraint}) induce perfect excitation transfer on even faster time scales (recall 
our discussion in Sec.~\ref{sec:desprinc} above).

After this first coarse-grained assessment of the statistics of the transfer efficiency via the competition between the 
scaled channel coupling and the distribution of disorder-induced relative level shifts of the doublet states, we now directly 
inspect the distribution of the transfer probabilities. To sample the latter at the dominant doublet energies $E' \pm V + s^{\pm}$ 
of the closed system \cite{walschaers_optimally_2013}, which follow from (\ref{eq:PResNLevel}) with $\Gamma =0$, and are slightly detuned with respect 
to (\ref{eq:PResNLevel}) for finite $\Gamma$, one needs to infer the statistics of 
\begin{equation}
p(E' \pm V+s^{\pm}) \approx \frac{(1+\Delta \tilde{s})^2}{(1+\Delta \tilde{s})^2+\tilde{\Gamma}^2/4} \label{eq:approxEnergiesP}\, ,
\end{equation}
as inherited from (\ref{eq:Cauchy}). This leads to 
\begin{equation}\label{eq:PpScatter}
P( p) = \int_{\mathbb{R}} {\rm d}\Delta\tilde{s}\, P(\Delta\tilde{s}) \delta\left(p - \frac{(1+\Delta\tilde{s})^2}{(1+\Delta\tilde{s})^2+\tilde{\Gamma}^2/4}\right)\, ,
\end{equation}
what can be evaluated with the help of
\begin{equation}
\delta(f(x))=\sum^m_{j=1} \frac{\delta(x-x_j)}{\abs{f'(x_j)}}
\end{equation} 
where $f(x_j)=0$ and $f'(x_j) \neq 0$. 
This 
in 
(\ref{eq:PpScatter}) yields
\begin{equation}\begin{split}
P(p)=&\frac{\tilde{\Gamma}}{4\pi\sqrt{p(1-p)^{3}}}
\\&\times\Bigg(\frac{\tilde{\sigma}}{\tilde{\sigma}^2+\left(1+\tilde{s}_0-\frac{\tilde{\Gamma}}{2}\sqrt{\frac{p}{1-p}}\right)^2}\\&\qquad+\frac{\tilde{\sigma}}{\tilde{\sigma}^2+\left(1+\tilde{s}_0+\frac{\tilde{\Gamma}}{2}\sqrt{\frac{p}{1-p}}\right)^2}\Bigg).
\label{eq:effDist}
\end{split}
\end{equation}
The condition $f'(x_j) \neq 0$ is violated on the edges of the 
domain $[0,1]$ 
of the probability distribution $P(p)$. 
The resulting distribution (\ref{eq:effDist}) is, therefore, only well-defined for $p \in [\delta, 1 - \delta]$. The divergences on the edges are an artefact of 
the power-law statistics (\ref{eq:Cauchy}) 
of $\Delta \tilde{s}$.

\begin{figure}[t]
\centering
\includegraphics[width=0.49\textwidth]{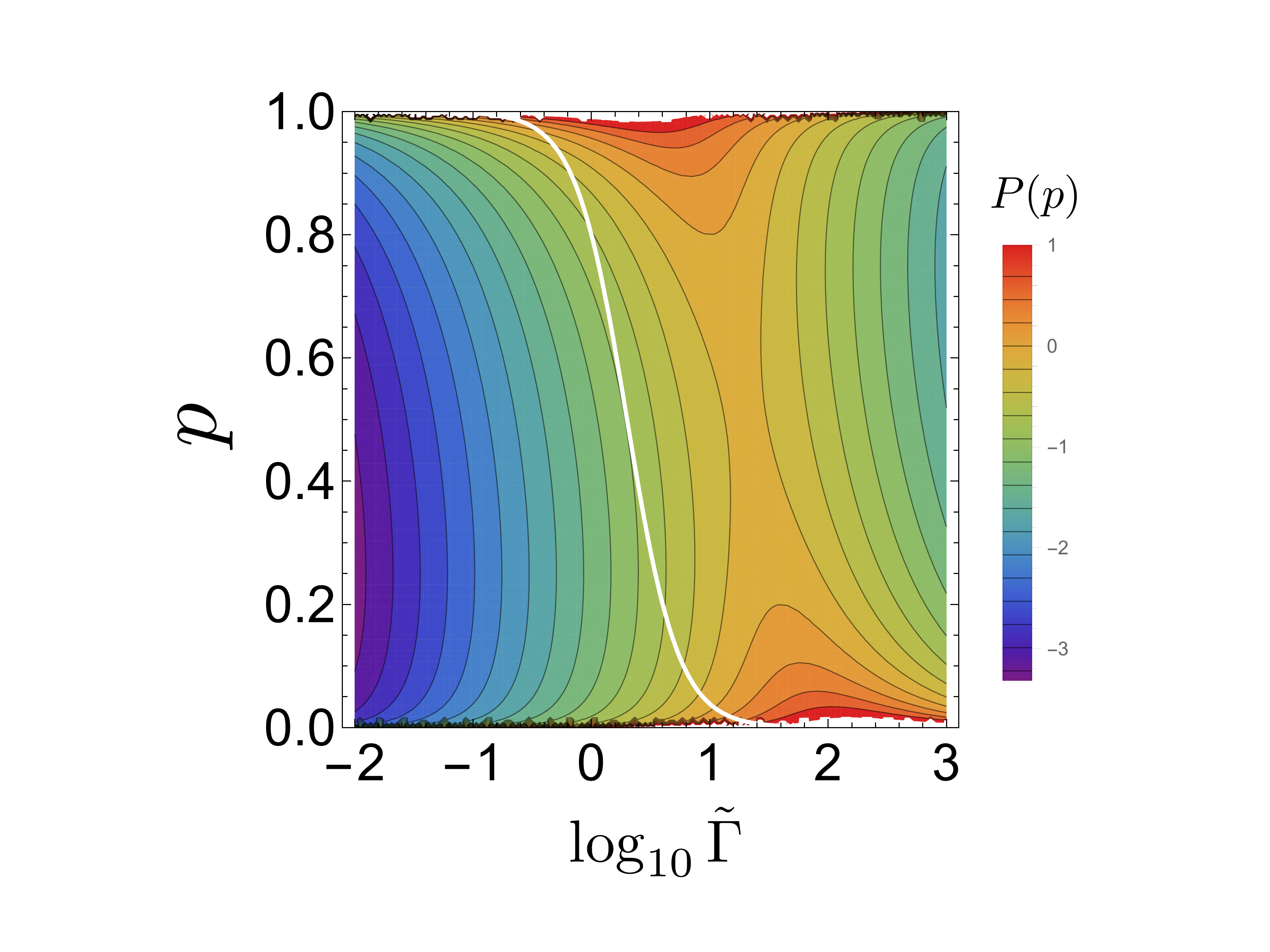}
\caption{Density plot of the probability density $P(p)$ (\ref{eq:effDist}) of the transfer efficiency as a function of $\tilde{\Gamma}$, for $\tilde{\sigma}=10$. Consistent with the dominant doublet condition (\ref{eq:doubletCond}, \ref{eq:domdubstat}), $\tilde{s}_0$ is set to zero. The white curve indicates the transfer probability for the two-level system without intermediate sites, as obtained by setting $s^+=s^-=0$ in (\ref{eq:approxEnergiesP}). Figure obtained from \cite{Walschaers_Thesis}.}
\label{fig:ProbDensP}
\end{figure}

In Fig.~\ref{fig:ProbDensP}, we compare  
the prediction of 
(\ref{eq:effDist}) (colour coded) 
to the transfer probability of a dimer (white line), which is obtained from (\ref{eq:HCentro}) with all couplings $v_i=0$ between input/output and 
bulk sites, and also implies $s^+=s^-=0$. For the dimer, it is clear from our previous discussion, as well from its very structure which is determined
by only two relevant
coupling constants, $V$ an $\Gamma$, that the fastest transfer time scales are achieved for $\tilde{\Gamma} \approx 1$, as clearly displayed by the plot. 
In contrast, in the presence of 
bulk sites, and for a broad distribution of the relative doublet shifts as assumed in Fig.~\ref{fig:ProbDensP} by the choice $\tilde{\sigma}=10$, the
transition between efficient and inefficient transport is pushed to values of $\tilde{\Gamma}$ enhanced by roughly one order of magnitude, an observation 
fully consistent with Fig.~\ref{fig:densEffReal}. 

\begin{figure*}
  \centering
\includegraphics[width=0.75\textwidth]{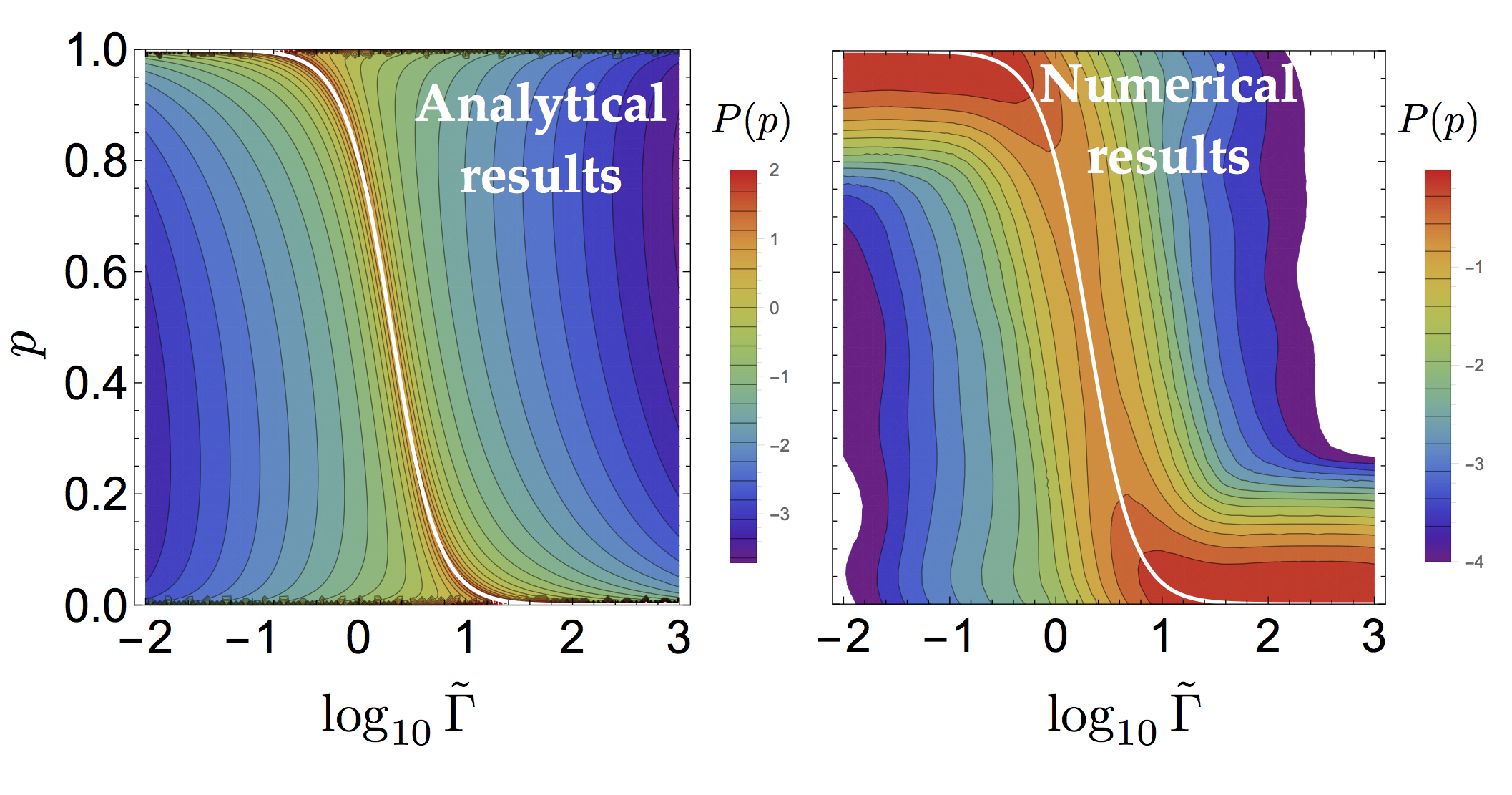}
\includegraphics[width=0.8\textwidth]{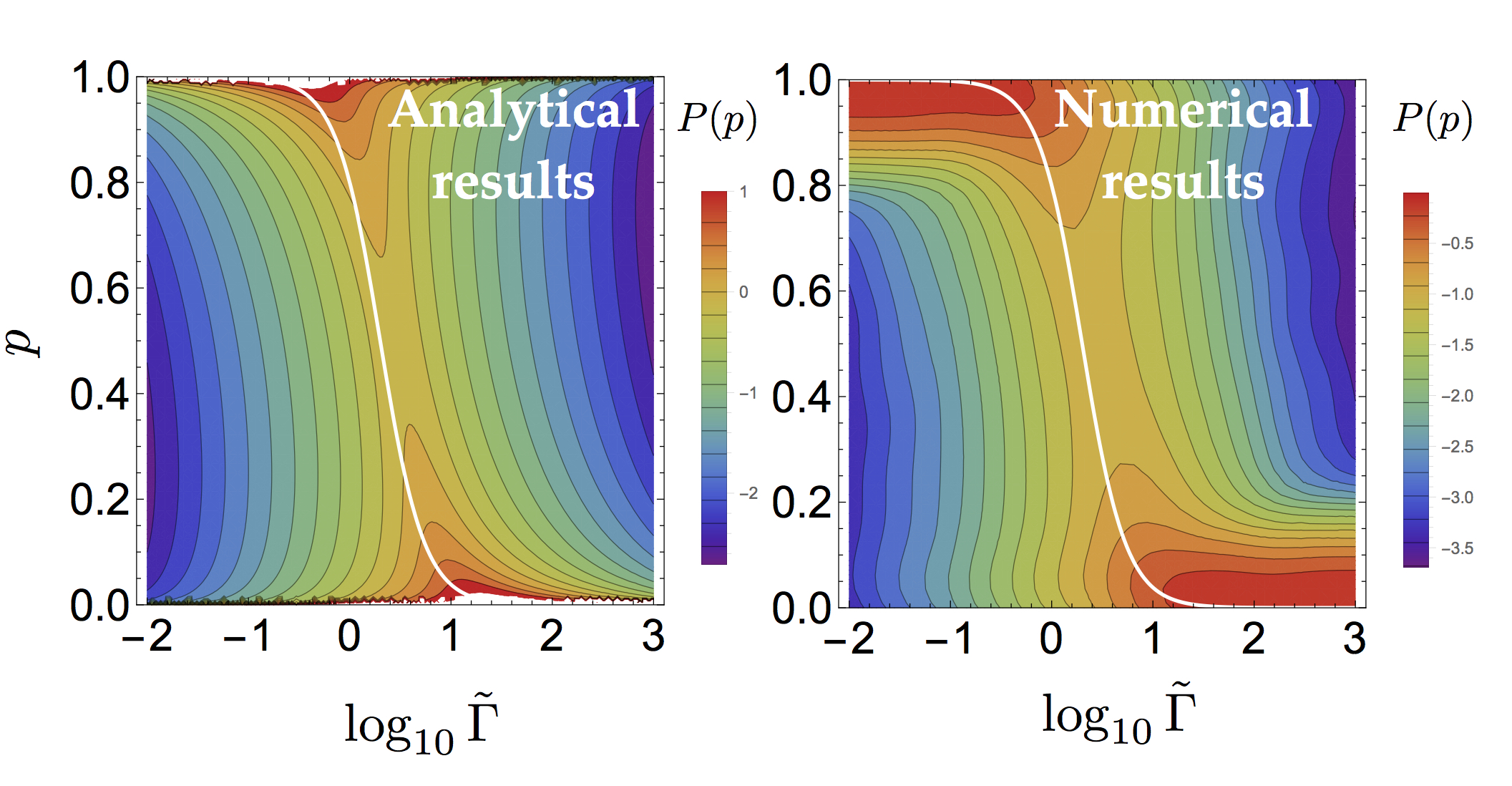}
\includegraphics[width=0.8\textwidth]{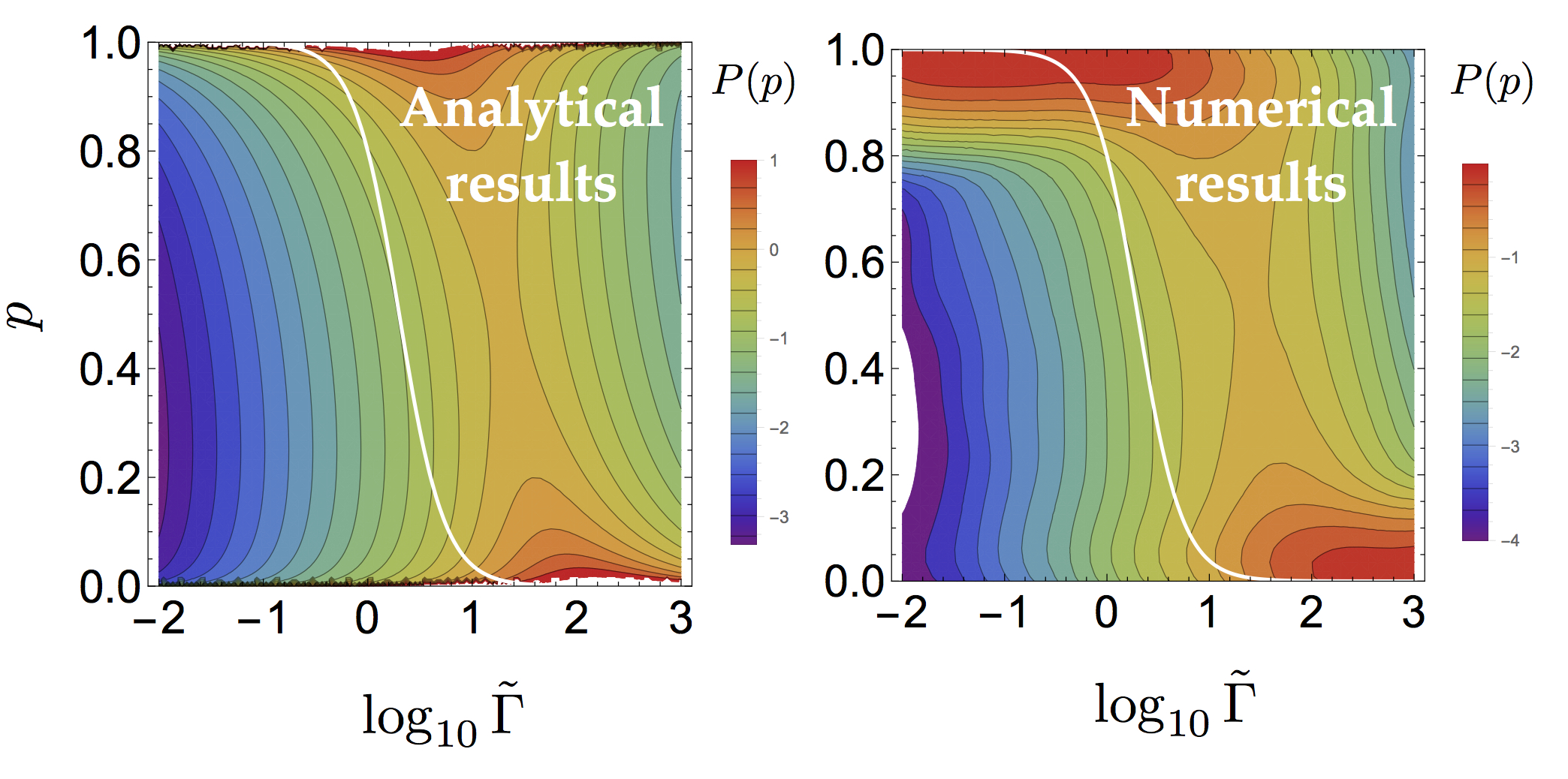}
\caption{Probability density $P(p)$ of transfer efficiencies (as density plot), as given by Eq.~(\ref{eq:effDist}) (left) and as obtained from numerics (right), for variable $\tilde{\Gamma}$. Different values for $\tilde{\sigma}$ (\ref{eq:sigmaands}) are shown in different rows: $\tilde{\sigma}=0.1$ (top) for eight sites with typical intermediate site coupling (\ref{eq:hierstaatdexi}) $\xi=20$; $\tilde{\sigma}=1$ (middle) for eight sites with $\xi = 50$; $\tilde{\sigma}=10$ (bottom) for ten sites with $\xi=150$. The dominant doublet constraint (\ref{eq:doubletCond}, \ref{eq:domdubstat}) is set to $\epsilon = 0.05$, which implicitly fixes all remaining parameters. The white curve indicates the transfer probability for the two-level system without intermediate sites, as obtained by setting $s^+=s^-=0$ in (\ref{eq:approxEnergiesP}). Figure obtained from \cite{Walschaers_Thesis}.}
\label{fig:ProbDensPTheoExp}
\end{figure*}

\section{Comparison to Numerical Results}\label{sec:NumericsForScattering}

Let us finally confront our analytical -- though perturbative (at lowest order) -- prediction (\ref{eq:PpScatter}) with model 
calculations for open, disordered finite networks. We follow the modelling in \cite{tomsovic_chaos-assisted_1994,leyvraz_level_1996}, and set
\begin{align}
&(H_{\rm int})_{i,j} \sim {\rm Normal}\left(0, (1+\delta_{i,j})\frac{\xi^2}{N}\right),\label{eq:hierstaatdexi}\\
&v_{i} \sim {\rm Normal}\left(0, \frac{\chi^2}{N}\right), \label{eq:hierstaatdechi}\\
&E' \sim   {\rm Normal}\left(0, 2\frac{\xi^2}{N}\right)\, ,
\end{align}
where $\chi$ and $\xi$ parametrize the typical (root mean square) values of the model Hamiltonian's (\ref{eq:HCentro}) stochastically distributed
coupling constants, and (\ref{eq:hierstaatdexi}) implies that the bulk sites' Hamiltonian $H_{\rm int}$ is sampled from the Gaussian orthogonal 
ensemble (GOE).\footnote{More details on this model can be found in Chapter 4 of \cite{Walschaers_Thesis}.}
We now combine the results of \cite{leyvraz_level_1996, ergun_level_2003} with the fact that for the intermediate sites the mean-level spacing is given by $\Delta = \pi \xi/(N/2-1)$  \cite{brouwer_random-matrix_1997}, and obtain that
\begin{equation}
\label{eq:sigmaands}
\tilde{\sigma} = \frac{\chi^2}{V\xi}, \quad \text{and} \quad \tilde{s}_0 = \frac{\chi^2}{2\xi^2}\, .
\end{equation}
Furthermore, 
the results in \cite{walschaers_statistical_2015, bohigas_manifestations_1993}, 
imply that, to fulfil the dominant doublet condition (\ref{eq:doubletCond}), we must have 
\begin{equation}\label{eq:domdubstat}
\left(\frac{2}{\pi}\right)^{3/2}\sqrt{\frac{N}{2}-1} \frac{\chi}{\xi} < \epsilon\, ,
\end{equation}
what ensures that 
$\tilde{s}_0$ in (\ref{eq:sigmaands}) is negligibly small, as already assumed above.

\begin{figure}
\centering
\includegraphics[width=0.49\textwidth]{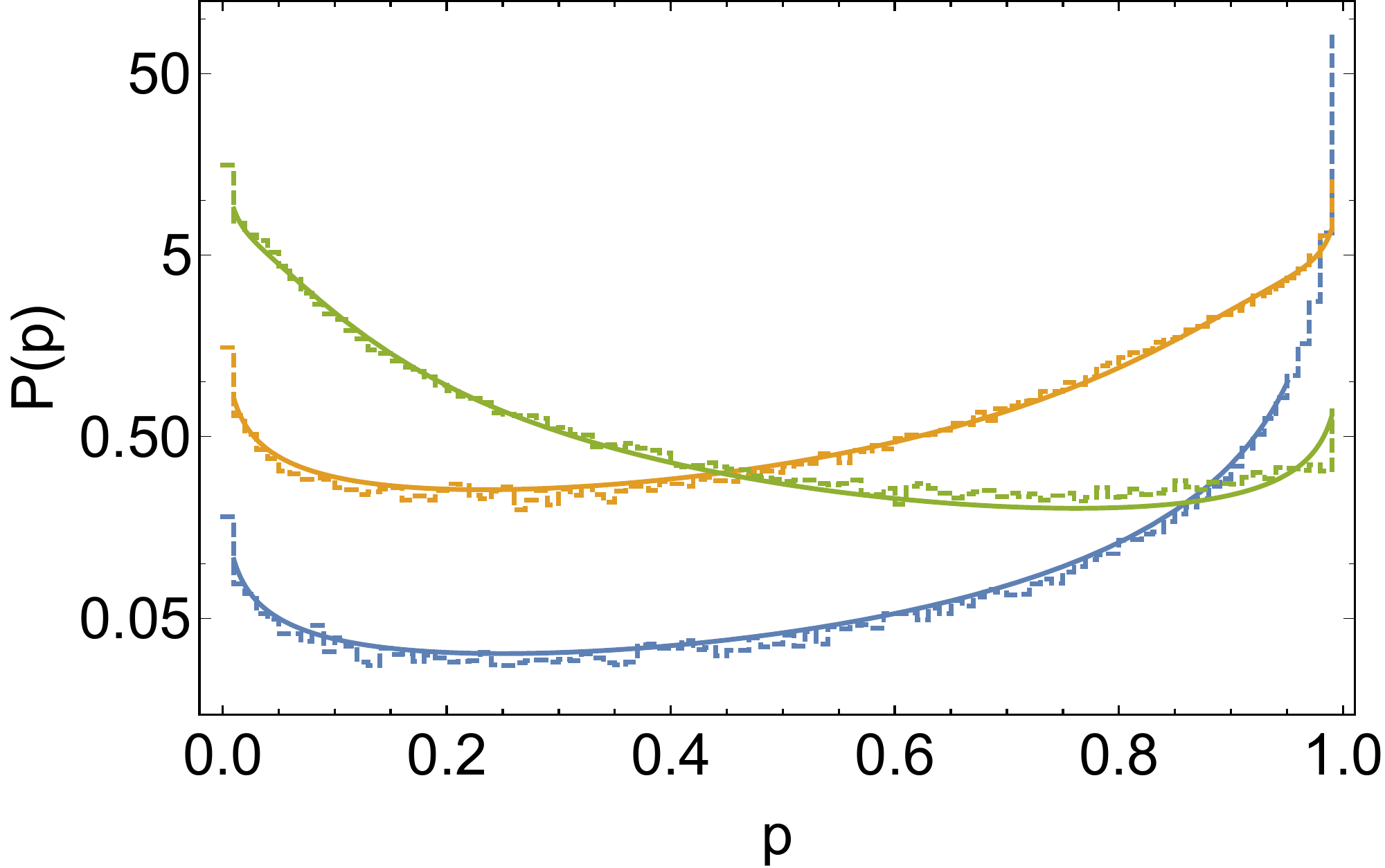}
\includegraphics[width=0.49\textwidth]{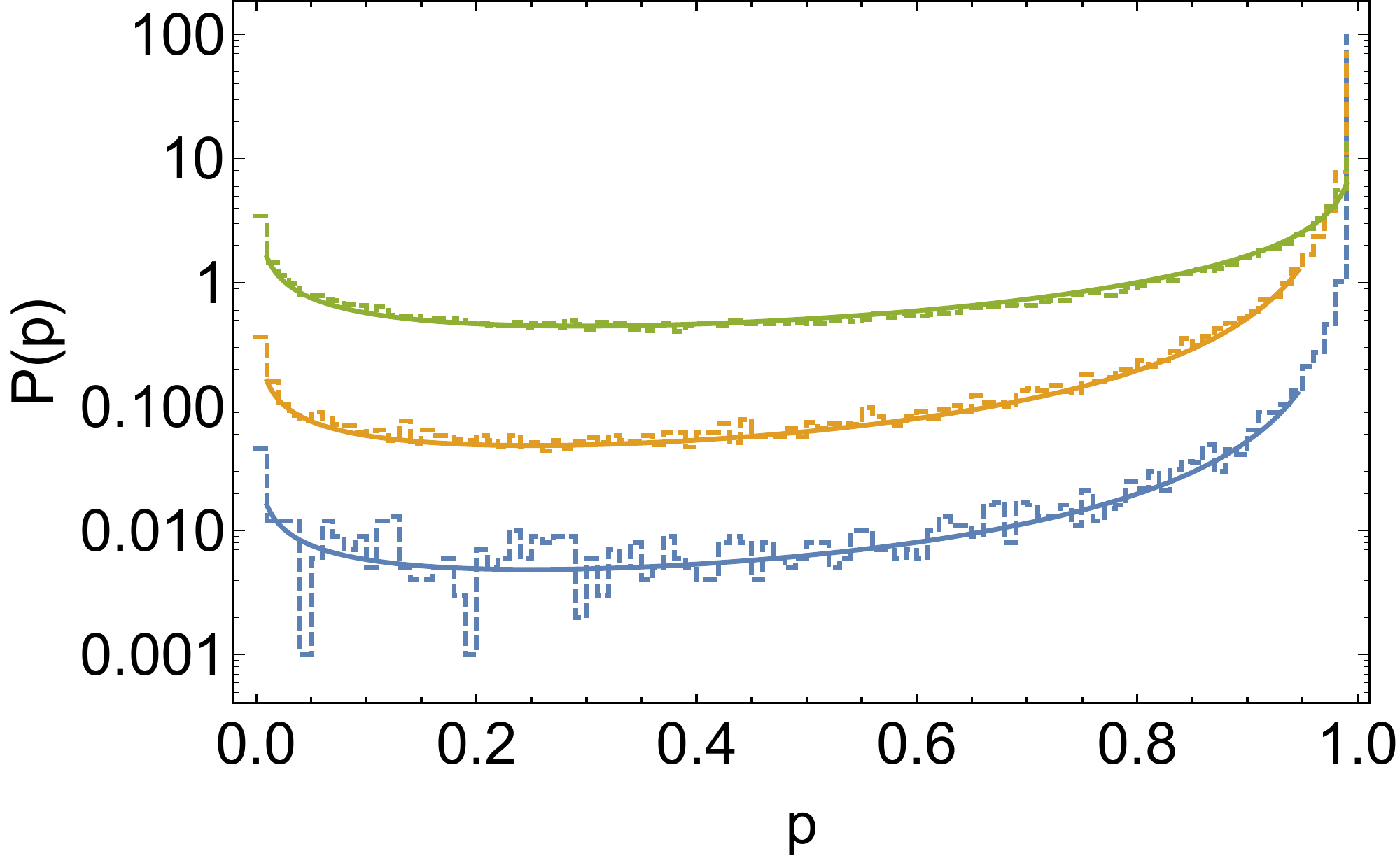}
\caption{Probability distribution for transfer probabilities as predicted theoretically by Eq.~(\ref{eq:effDist}) (solid lines), and as obtained (histograms, dashed lines) from numerical simulations. The width $\tilde{\sigma}$ of the relative shift distribution (\ref{eq:Cauchy}), was chosen as $\tilde{\sigma} = 1$ (top) and $\tilde{\sigma} = 10$ (bottom). In each panel results are shown for three different values of the rescaled channel coupling strength: $\tilde{\Gamma}=0.1$ (blue), $\tilde{\Gamma}=1$ (orange) and $\tilde{\Gamma}=10$ (green). Figure obtained from \cite{Walschaers_Thesis}.
}
\label{fig:ExtractionProbDensPTheoExp}
\end{figure}

In our simulations, we explicitly evaluate the scattering matrix element (\ref{eq:Smatrix}) and the transfer efficiency (\ref{eq:PScatter}) from 
the input to the output channel, at energy $E=E'+V+s^+$. $N$, $V$, $\xi$, and $\chi$ are all varied in the three numerical simulations displayed in
Fig.~\ref{fig:ProbDensPTheoExp}, to broadly cover parameter space.
To generate data, we choose different values for $\Gamma$, adopted to logarithmic scaling of the data represented in Figs.~\ref{fig:ProbDensPTheoExp} 
and \ref{fig:ExtractionProbDensPTheoExp} (for practical purposes we chose $\Gamma_{i+1} = 1.2\, \Gamma_{i}$). For each 
$\Gamma$, $10000$ random Hamiltonians (\ref{eq:HCentro}) with (\ref{eq:hierstaatdexi}) are diagonalized to extract 
the shift $s^+$. 
This allows the ultimate evaluation of 
$S_{{\rm in} \rightarrow {\rm out}}(E'+V+s^+)$,
and Fig.~\ref{fig:ProbDensPTheoExp} is output by the 
extrapolation method $\mathtt{SmoothKernelDistribution}$ of Mathematica, with the numerically obtained, discrete data points as input. 

Fig.~\ref{fig:ProbDensPTheoExp} shows good qualitative agreement between the theoretically obtained probability distribution (\ref{eq:effDist}) and the numerical results, what confirms that 
$\tilde{\sigma}$ and $\tilde{\Gamma}$ ultimately are the only 
parameters 
which control the statistics of transfer efficiencies. This physically implies that it is not the absolute energy scales
set by $\chi$ and $\xi$ 
in (\ref{eq:hierstaatdexi}) and (\ref{eq:hierstaatdechi}), respectively, but rather the {\em ratio} of the coupling strengths (i.e., $\tilde{\sigma}$ 
and $\tilde{\Gamma}$) which govern the efficiency of the transfer process. The only absolute energy scale to affect the physics is the 
coupling $\Gamma$ to the external channels, because it directly controls the time scale of the transport.

Inspecting Fig.~\ref{fig:ProbDensPTheoExp} in more detail does also reveal some quantitative differences between simulation data and theory.
Several features, even though they 
emerge at the same values of $\tilde{\Gamma}$, are considerably broader in the numerical data.
To verify whether these discrepancies are numerical artefacts induced by the here employed extrapolation method or due to 
essential physical features which are missed by our analytical treatment, we select three values of $\tilde{\Gamma}$, of different orders of 
magnitude, and compare 
analytical prediction (\ref{eq:PpScatter}) and numerical results for the density $P(p)$ in Fig.~\ref{fig:ProbDensPTheoExp}. 
Clearly, the 
comparison is quantitatively excellent in all dynamical regimes (defined by the different values of $\tilde{\Gamma}$), what identifies the above 
discrepancies
as a deficiency of the extrapolation method employed to produce Fig.~\ref{fig:ProbDensPTheoExp}. More importantly, this comparison also 
shows that our analysis, and (\ref{eq:densEffReal}) and (\ref{eq:PpScatter}) as the central results thereof, provide statistical guidance to optimise 
the transfer efficiencies of disordered networks, ultimately through $\tilde{\sigma}$ 
and $\tilde{\Gamma}$ alone.

\section{Conclusions}

Our present contribution provides a scattering theoretical generalisation of earlier results \cite{weber_transport_2010,zech_centrosymmetry_2014,levi_quantum_2015,scholak_efficient_2011,ortega_quantum_2015,PhysRevE.94.042102} 
on optimal excitation transfer across 
finite disordered networks of dipole coupled two-level systems. Given disordered networks which exhibit the design principles
already identified for closed systems -- centrosymmetry and a dominant doublet -- we have shown that, in order to achieve near-to-complete 
excitation transfer across such a network in minimal time, the coupling strength to the leads needs to be as large as possible, yet 
smaller than the dominant doublet splitting, as expressed by (\ref{eq:constraint}) above. In other words, the period of the 
coherent oscillation of the excitation between input and output site, as essentially defined by the dominant doublet splitting in the closed model 
\cite{walschaers_optimally_2013,walschaers_statistical_2015}, has to be shorter than the decay time of the resonances associated with 
the doublet states. Consequently, correlation functions which quantify, e.g., the population of the input or of the output site, must exhibit (rapidly decaying)
oscillations, as also familiar from typical 2D data recorded in experiments on photosynthetic light harvesting units \cite{engel_evidence_2007}. Our results
show in a rather transparent way that such separation of time scales, expressed by (\ref{eq:constraint}) and recently evidenced in experiments \cite{FMO_NatChem}
on the molecular superstructure the FMO complex is embedded in, is a necessary condition for faithful transfer
of the excitation: Increasing the channel coupling beyond the doublet splitting unavoidably reduces the transfer probability. 

We have further 
shown that disorder-induced fluctuations of the doublet splitting as brought about by slow drifts and/or random distributions of the microscopic 
network hardwiring, and observed e.g. in the distribution of B800-B850 coupling strengths in experiments on the LH2 complex 
\cite{hildner_quantum_2013}, 
allow to accelerate the transfer process under the above optimality condition (\ref{eq:constraint}): A broad distribution of 
relative shifts of the doublet levels allows for enhanced channel couplings, and, hence, for accelerated transmission,
which would lead to a net loss of population transfer in the absence of disorder. Note that the same fluctuations also lead to an effective broadening
of the condition for resonant excitation transfer when measured on a disordered ensemble.

\section*{Acknowledgements}
M.W. thanks the German National Academic Foundation for financial support. R.M. is indebted to the Alexander von Humboldt foundation 
for support through a research fellowship. A.B. acknowledges funding through EU Collaborative project QuProCS (Grant 
Agreement No. 641277).

\bibliography{Bib_Thesis_Update}

\begin{thebibliography}{47}%
\makeatletter
\providecommand \@ifxundefined [1]{%
 \@ifx{#1\undefined}
}%
\providecommand \@ifnum [1]{%
 \ifnum #1\expandafter \@firstoftwo
 \else \expandafter \@secondoftwo
 \fi
}%
\providecommand \@ifx [1]{%
 \ifx #1\expandafter \@firstoftwo
 \else \expandafter \@secondoftwo
 \fi
}%
\providecommand \natexlab [1]{#1}%
\providecommand \enquote  [1]{``#1''}%
\providecommand \bibnamefont  [1]{#1}%
\providecommand \bibfnamefont [1]{#1}%
\providecommand \citenamefont [1]{#1}%
\providecommand \href@noop [0]{\@secondoftwo}%
\providecommand \href [0]{\begingroup \@sanitize@url \@href}%
\providecommand \@href[1]{\@@startlink{#1}\@@href}%
\providecommand \@@href[1]{\endgroup#1\@@endlink}%
\providecommand \@sanitize@url [0]{\catcode `\\12\catcode `\$12\catcode
  `\&12\catcode `\#12\catcode `\^12\catcode `\_12\catcode `\%12\relax}%
\providecommand \@@startlink[1]{}%
\providecommand \@@endlink[0]{}%
\providecommand \url  [0]{\begingroup\@sanitize@url \@url }%
\providecommand \@url [1]{\endgroup\@href {#1}{\urlprefix }}%
\providecommand \urlprefix  [0]{URL }%
\providecommand \Eprint [0]{\href }%
\providecommand \doibase [0]{http://dx.doi.org/}%
\providecommand \selectlanguage [0]{\@gobble}%
\providecommand \bibinfo  [0]{\@secondoftwo}%
\providecommand \bibfield  [0]{\@secondoftwo}%
\providecommand \translation [1]{[#1]}%
\providecommand \BibitemOpen [0]{}%
\providecommand \bibitemStop [0]{}%
\providecommand \bibitemNoStop [0]{.\EOS\space}%
\providecommand \EOS [0]{\spacefactor3000\relax}%
\providecommand \BibitemShut  [1]{\csname bibitem#1\endcsname}%
\let\auto@bib@innerbib\@empty
\bibitem [{\citenamefont {Scholak}\ \emph {et~al.}(2010)\citenamefont
  {Scholak}, \citenamefont {Mintert}, \citenamefont {Wellens},\ and\
  \citenamefont {Buchleitner}}]{weber_transport_2010}%
  \BibitemOpen
  \bibfield  {author} {\bibinfo {author} {\bibfnamefont {T.}~\bibnamefont
  {Scholak}}, \bibinfo {author} {\bibfnamefont {F.}~\bibnamefont {Mintert}},
  \bibinfo {author} {\bibfnamefont {T.}~\bibnamefont {Wellens}}, \ and\
  \bibinfo {author} {\bibfnamefont {A.}~\bibnamefont {Buchleitner}},\ }in\
  \href@noop {} {{\selectlanguage {eng}\emph {\bibinfo {booktitle}
  {Biomolecular systems}}}},\ \bibinfo {series and number} {Quantum efficiency
  in complex systems},\ \bibinfo {editor} {edited by\ \bibinfo {editor}
  {\bibfnamefont {E.~R.}\ \bibnamefont {Weber}}, \bibinfo {editor}
  {\bibfnamefont {M.}~\bibnamefont {Thorwart}}, \ and\ \bibinfo {editor}
  {\bibfnamefont {U.}~\bibnamefont {W{\"u}rfel}}}\ (\bibinfo  {publisher}
  {Elsevier},\ \bibinfo {address} {Oxford},\ \bibinfo {year} {2010})\ \bibinfo
  {edition} {1st}\ ed.\BibitemShut {Stop}%
\bibitem [{\citenamefont {Levi}\ \emph {et~al.}(2015)\citenamefont {Levi},
  \citenamefont {Mostarda}, \citenamefont {Rao},\ and\ \citenamefont
  {Mintert}}]{levi_quantum_2015}%
  \BibitemOpen
  \bibfield  {author} {\bibinfo {author} {\bibfnamefont {F.}~\bibnamefont
  {Levi}}, \bibinfo {author} {\bibfnamefont {S.}~\bibnamefont {Mostarda}},
  \bibinfo {author} {\bibfnamefont {F.}~\bibnamefont {Rao}}, \ and\ \bibinfo
  {author} {\bibfnamefont {F.}~\bibnamefont {Mintert}},\ }\href {\doibase
  10.1088/0034-4885/78/8/082001} {\bibfield  {journal} {\bibinfo  {journal}
  {Rep. Prog. Phys.}\ }\textbf {\bibinfo {volume} {78}},\ \bibinfo {pages}
  {082001} (\bibinfo {year} {2015})}\BibitemShut {NoStop}%
\bibitem [{\citenamefont {Walschaers}\ \emph {et~al.}(2016)\citenamefont
  {Walschaers}, \citenamefont {Schlawin}, \citenamefont {Wellens},\ and\
  \citenamefont {Buchleitner}}]{doi:10.1146/annurev-conmatphys-031115-011327}%
  \BibitemOpen
  \bibfield  {author} {\bibinfo {author} {\bibfnamefont {M.}~\bibnamefont
  {Walschaers}}, \bibinfo {author} {\bibfnamefont {F.}~\bibnamefont
  {Schlawin}}, \bibinfo {author} {\bibfnamefont {T.}~\bibnamefont {Wellens}}, \
  and\ \bibinfo {author} {\bibfnamefont {A.}~\bibnamefont {Buchleitner}},\
  }\href {\doibase 10.1146/annurev-conmatphys-031115-011327} {\bibfield
  {journal} {\bibinfo  {journal} {Annual Review of Condensed Matter Physics}\
  }\textbf {\bibinfo {volume} {7}},\ \bibinfo {pages} {223} (\bibinfo {year}
  {2016})},\ \Eprint
  {http://arxiv.org/abs/http://dx.doi.org/10.1146/annurev-conmatphys-031115-011327}
  {http://dx.doi.org/10.1146/annurev-conmatphys-031115-011327} \BibitemShut
  {NoStop}%
\bibitem [{\citenamefont {Spallek~et al.}(2017)}]{Spallek}%
  \BibitemOpen
  \bibfield  {author} {\bibinfo {author} {\bibfnamefont {F.}~\bibnamefont
  {Spallek~et al.}},\ }\href@noop {} {\bibfield  {journal} {\bibinfo  {journal}
  {submitted to J. Phys. B, this Special Issue}\ } (\bibinfo {year}
  {2017})}\BibitemShut {NoStop}%
\bibitem [{\citenamefont {Gurian}\ \emph {et~al.}(2012)\citenamefont {Gurian},
  \citenamefont {Cheinet}, \citenamefont {Huillery}, \citenamefont {Fioretti},
  \citenamefont {Zhao}, \citenamefont {Gould}, \citenamefont {Comparat},\ and\
  \citenamefont {Pillet}}]{PhysRevLett.108.023005}%
  \BibitemOpen
  \bibfield  {author} {\bibinfo {author} {\bibfnamefont {J.~H.}\ \bibnamefont
  {Gurian}}, \bibinfo {author} {\bibfnamefont {P.}~\bibnamefont {Cheinet}},
  \bibinfo {author} {\bibfnamefont {P.}~\bibnamefont {Huillery}}, \bibinfo
  {author} {\bibfnamefont {A.}~\bibnamefont {Fioretti}}, \bibinfo {author}
  {\bibfnamefont {J.}~\bibnamefont {Zhao}}, \bibinfo {author} {\bibfnamefont
  {P.~L.}\ \bibnamefont {Gould}}, \bibinfo {author} {\bibfnamefont
  {D.}~\bibnamefont {Comparat}}, \ and\ \bibinfo {author} {\bibfnamefont
  {P.}~\bibnamefont {Pillet}},\ }\href {\doibase
  10.1103/PhysRevLett.108.023005} {\bibfield  {journal} {\bibinfo  {journal}
  {Phys. Rev. Lett.}\ }\textbf {\bibinfo {volume} {108}},\ \bibinfo {pages}
  {023005} (\bibinfo {year} {2012})}\BibitemShut {NoStop}%
\bibitem [{\citenamefont {Scholak}\ \emph {et~al.}(2014)\citenamefont
  {Scholak}, \citenamefont {Wellens},\ and\ \citenamefont
  {Buchleitner}}]{scholak_spectral_2014}%
  \BibitemOpen
  \bibfield  {author} {\bibinfo {author} {\bibfnamefont {T.}~\bibnamefont
  {Scholak}}, \bibinfo {author} {\bibfnamefont {T.}~\bibnamefont {Wellens}}, \
  and\ \bibinfo {author} {\bibfnamefont {A.}~\bibnamefont {Buchleitner}},\
  }\href {\doibase 10.1103/PhysRevA.90.063415} {\bibfield  {journal} {\bibinfo
  {journal} {Phys. Rev. A}\ }\textbf {\bibinfo {volume} {90}},\ \bibinfo
  {pages} {063415} (\bibinfo {year} {2014})}\BibitemShut {NoStop}%
\bibitem [{\citenamefont {Deng}\ \emph {et~al.}(2016)\citenamefont {Deng},
  \citenamefont {Altshuler}, \citenamefont {Shlyapnikov},\ and\ \citenamefont
  {Santos}}]{Deng:2016aa}%
  \BibitemOpen
  \bibfield  {author} {\bibinfo {author} {\bibfnamefont {X.}~\bibnamefont
  {Deng}}, \bibinfo {author} {\bibfnamefont {B.~L.}\ \bibnamefont {Altshuler}},
  \bibinfo {author} {\bibfnamefont {G.~V.}\ \bibnamefont {Shlyapnikov}}, \ and\
  \bibinfo {author} {\bibfnamefont {L.}~\bibnamefont {Santos}},\ }\href
  {https://link.aps.org/doi/10.1103/PhysRevLett.117.020401} {\bibfield
  {journal} {\bibinfo  {journal} {Physical Review Letters}\ }\textbf {\bibinfo
  {volume} {117}},\ \bibinfo {pages} {020401} (\bibinfo {year}
  {2016})}\BibitemShut {NoStop}%
\bibitem [{\citenamefont {Dost{\'a}l}\ \emph {et~al.}(2016)\citenamefont
  {Dost{\'a}l}, \citenamefont {P{\v s}en{\v c}{\'\i}k},\ and\ \citenamefont
  {Zigmantas}}]{FMO_NatChem}%
  \BibitemOpen
  \bibfield  {author} {\bibinfo {author} {\bibfnamefont {J.}~\bibnamefont
  {Dost{\'a}l}}, \bibinfo {author} {\bibfnamefont {J.}~\bibnamefont {P{\v
  s}en{\v c}{\'\i}k}}, \ and\ \bibinfo {author} {\bibfnamefont
  {D.}~\bibnamefont {Zigmantas}},\ }\href
  {http://dx.doi.org/10.1038/nchem.2525} {\bibfield  {journal} {\bibinfo
  {journal} {Nat Chem}\ }\textbf {\bibinfo {volume} {8}},\ \bibinfo {pages}
  {705} (\bibinfo {year} {2016})}\BibitemShut {NoStop}%
\bibitem [{\citenamefont {Kramer}\ and\ \citenamefont
  {Rodriguez}(2017)}]{Kramer17}%
  \BibitemOpen
  \bibfield  {author} {\bibinfo {author} {\bibfnamefont {T.}~\bibnamefont
  {Kramer}}\ and\ \bibinfo {author} {\bibfnamefont {M.}~\bibnamefont
  {Rodriguez}},\ }\href {http://dx.doi.org/10.1038/srep45245} {\bibfield
  {journal} {\bibinfo  {journal} {Scientific Reports}\ }\textbf {\bibinfo
  {volume} {7}},\ \bibinfo {pages} {45245 EP } (\bibinfo {year}
  {2017})}\BibitemShut {NoStop}%
\bibitem [{\citenamefont {Ringsmuth}\ \emph {et~al.}(2012)\citenamefont
  {Ringsmuth}, \citenamefont {Milburn},\ and\ \citenamefont
  {Stace}}]{Ringsmuth2012}%
  \BibitemOpen
  \bibfield  {author} {\bibinfo {author} {\bibfnamefont {A.~K.}\ \bibnamefont
  {Ringsmuth}}, \bibinfo {author} {\bibfnamefont {G.~J.}\ \bibnamefont
  {Milburn}}, \ and\ \bibinfo {author} {\bibfnamefont {T.~M.}\ \bibnamefont
  {Stace}},\ }\href {http://dx.doi.org/10.1038/nphys2332} {\bibfield  {journal}
  {\bibinfo  {journal} {Nat Phys}\ }\textbf {\bibinfo {volume} {8}},\ \bibinfo
  {pages} {562} (\bibinfo {year} {2012})}\BibitemShut {NoStop}%
\bibitem [{\citenamefont {Fano}(1961)}]{fano_effects_1961}%
  \BibitemOpen
  \bibfield  {author} {\bibinfo {author} {\bibfnamefont {U.}~\bibnamefont
  {Fano}},\ }\href {\doibase 10.1103/PhysRev.124.1866} {\bibfield  {journal}
  {\bibinfo  {journal} {Phys. Rev.}\ }\textbf {\bibinfo {volume} {124}},\
  \bibinfo {pages} {1866} (\bibinfo {year} {1961})}\BibitemShut {NoStop}%
\bibitem [{\citenamefont {Feshbach}(1958)}]{feshbach_unified_1958}%
  \BibitemOpen
  \bibfield  {author} {\bibinfo {author} {\bibfnamefont {H.}~\bibnamefont
  {Feshbach}},\ }\href {\doibase 10.1016/0003-4916(58)90007-1} {\bibfield
  {journal} {\bibinfo  {journal} {Ann. Phys.}\ }\textbf {\bibinfo {volume}
  {5}},\ \bibinfo {pages} {357} (\bibinfo {year} {1958})}\BibitemShut {NoStop}%
\bibitem [{\citenamefont {Feshbach}(1962)}]{feshbach_unified_1962}%
  \BibitemOpen
  \bibfield  {author} {\bibinfo {author} {\bibfnamefont {H.}~\bibnamefont
  {Feshbach}},\ }\href {\doibase 10.1016/0003-4916(62)90221-X} {\bibfield
  {journal} {\bibinfo  {journal} {Ann. Phys.}\ }\textbf {\bibinfo {volume}
  {19}},\ \bibinfo {pages} {287} (\bibinfo {year} {1962})}\BibitemShut
  {NoStop}%
\bibitem [{\citenamefont {Feshbach}(1967)}]{feshbach_unified_1967}%
  \BibitemOpen
  \bibfield  {author} {\bibinfo {author} {\bibfnamefont {H.}~\bibnamefont
  {Feshbach}},\ }\href {\doibase 10.1016/0003-4916(67)90163-7} {\bibfield
  {journal} {\bibinfo  {journal} {Ann. Phys.}\ }\textbf {\bibinfo {volume}
  {43}},\ \bibinfo {pages} {410} (\bibinfo {year} {1967})}\BibitemShut
  {NoStop}%
\bibitem [{Note1()}]{Note1}%
  \BibitemOpen
  \bibinfo {note} {Direct application of the formalism of \cite
  {feshbach_unified_1958,feshbach_unified_1962,feshbach_unified_1967} leads to
  $$S(E)=\protect \mathbb {1}- 2 \pi i \protect \mathaccentV
  {hat}05E{W}^{\protect \dag }\protect \frac {1}{E - H_{\protect \rm
  eff}}\protect \mathaccentV {hat}05E{W}\protect \tmspace +\thinmuskip
  {.1667em} ,$$ hence $W = \protect \sqrt {\pi }\protect \mathaccentV
  {hat}05E{W}.$ Both conventions appear in the literature, and we choose (\ref
  {eq:Smatrix}) for notational convenience.}\BibitemShut {Stop}%
\bibitem [{\citenamefont {Brouwer}\ \emph {et~al.}(1997)\citenamefont
  {Brouwer}, \citenamefont {Frahm},\ and\ \citenamefont
  {Beenakker}}]{brouwer_quantum_1997}%
  \BibitemOpen
  \bibfield  {author} {\bibinfo {author} {\bibfnamefont {P.~W.}\ \bibnamefont
  {Brouwer}}, \bibinfo {author} {\bibfnamefont {K.~M.}\ \bibnamefont {Frahm}},
  \ and\ \bibinfo {author} {\bibfnamefont {C.~W.~J.}\ \bibnamefont
  {Beenakker}},\ }\href {\doibase 10.1103/PhysRevLett.78.4737} {\bibfield
  {journal} {\bibinfo  {journal} {Phys. Rev. Lett.}\ }\textbf {\bibinfo
  {volume} {78}},\ \bibinfo {pages} {4737} (\bibinfo {year}
  {1997})}\BibitemShut {NoStop}%
\bibitem [{\citenamefont {Celardo}\ and\ \citenamefont
  {Kaplan}(2009)}]{celardo_superradiance_2009}%
  \BibitemOpen
  \bibfield  {author} {\bibinfo {author} {\bibfnamefont {G.~L.}\ \bibnamefont
  {Celardo}}\ and\ \bibinfo {author} {\bibfnamefont {L.}~\bibnamefont
  {Kaplan}},\ }\href {\doibase 10.1103/PhysRevB.79.155108} {\bibfield
  {journal} {\bibinfo  {journal} {Phys. Rev. B}\ }\textbf {\bibinfo {volume}
  {79}},\ \bibinfo {pages} {155108} (\bibinfo {year} {2009})}\BibitemShut
  {NoStop}%
\bibitem [{\citenamefont {Haake}\ \emph {et~al.}(1992)\citenamefont {Haake},
  \citenamefont {Izrailev}, \citenamefont {Lehmann}, \citenamefont {Saher},\
  and\ \citenamefont {Sommers}}]{haake_statistics_1992}%
  \BibitemOpen
  \bibfield  {author} {\bibinfo {author} {\bibfnamefont {F.}~\bibnamefont
  {Haake}}, \bibinfo {author} {\bibfnamefont {F.}~\bibnamefont {Izrailev}},
  \bibinfo {author} {\bibfnamefont {N.}~\bibnamefont {Lehmann}}, \bibinfo
  {author} {\bibfnamefont {D.}~\bibnamefont {Saher}}, \ and\ \bibinfo {author}
  {\bibfnamefont {H.-J.}\ \bibnamefont {Sommers}},\ }\href {\doibase
  10.1007/BF01470925} {\bibfield  {journal} {\bibinfo  {journal} {Z. Physik B -
  Condensed Matter}\ }\textbf {\bibinfo {volume} {88}},\ \bibinfo {pages} {359}
  (\bibinfo {year} {1992})}\BibitemShut {NoStop}%
\bibitem [{\citenamefont {Lewenkopf}\ and\ \citenamefont
  {Weidenm{\"u}ller}(1991)}]{lewenkopf_stochastic_1991}%
  \BibitemOpen
  \bibfield  {author} {\bibinfo {author} {\bibfnamefont {C.~H.}\ \bibnamefont
  {Lewenkopf}}\ and\ \bibinfo {author} {\bibfnamefont {H.~A.}\ \bibnamefont
  {Weidenm{\"u}ller}},\ }\href {\doibase 10.1016/0003-4916(91)90372-F}
  {\bibfield  {journal} {\bibinfo  {journal} {Ann. Phys.}\ }\textbf {\bibinfo
  {volume} {212}},\ \bibinfo {pages} {53} (\bibinfo {year} {1991})}\BibitemShut
  {NoStop}%
\bibitem [{\citenamefont {{\v S}eba}\ \emph {et~al.}(1996)\citenamefont {{\v
  S}eba}, \citenamefont {{\.Z}yczkowski},\ and\ \citenamefont
  {Zakrzewski}}]{seba_statistical_1996}%
  \BibitemOpen
  \bibfield  {author} {\bibinfo {author} {\bibfnamefont {P.}~\bibnamefont {{\v
  S}eba}}, \bibinfo {author} {\bibfnamefont {K.}~\bibnamefont
  {{\.Z}yczkowski}}, \ and\ \bibinfo {author} {\bibfnamefont {J.}~\bibnamefont
  {Zakrzewski}},\ }\href {\doibase 10.1103/PhysRevE.54.2438} {\bibfield
  {journal} {\bibinfo  {journal} {Phys. Rev. E}\ }\textbf {\bibinfo {volume}
  {54}},\ \bibinfo {pages} {2438} (\bibinfo {year} {1996})}\BibitemShut
  {NoStop}%
\bibitem [{\citenamefont {St{\"o}ckmann}\ \emph {et~al.}(2002)\citenamefont
  {St{\"o}ckmann}, \citenamefont {Persson}, \citenamefont {Kim}, \citenamefont
  {Barth}, \citenamefont {Kuhl},\ and\ \citenamefont
  {Rotter}}]{stockmann_effective_2002}%
  \BibitemOpen
  \bibfield  {author} {\bibinfo {author} {\bibfnamefont {H.-J.}\ \bibnamefont
  {St{\"o}ckmann}}, \bibinfo {author} {\bibfnamefont {E.}~\bibnamefont
  {Persson}}, \bibinfo {author} {\bibfnamefont {Y.-H.}\ \bibnamefont {Kim}},
  \bibinfo {author} {\bibfnamefont {M.}~\bibnamefont {Barth}}, \bibinfo
  {author} {\bibfnamefont {U.}~\bibnamefont {Kuhl}}, \ and\ \bibinfo {author}
  {\bibfnamefont {I.}~\bibnamefont {Rotter}},\ }\href {\doibase
  10.1103/PhysRevE.65.066211} {\bibfield  {journal} {\bibinfo  {journal} {Phys.
  Rev. E}\ }\textbf {\bibinfo {volume} {65}},\ \bibinfo {pages} {066211}
  (\bibinfo {year} {2002})}\BibitemShut {NoStop}%
\bibitem [{\citenamefont {Rotter}(2009)}]{rotter_non-hermitian_2009}%
  \BibitemOpen
  \bibfield  {author} {\bibinfo {author} {\bibfnamefont {I.}~\bibnamefont
  {Rotter}},\ }\href {\doibase 10.1088/1751-8113/42/15/153001} {\bibfield
  {journal} {\bibinfo  {journal} {J. Phys. A: Math. Theor.}\ }\textbf {\bibinfo
  {volume} {42}},\ \bibinfo {pages} {153001} (\bibinfo {year}
  {2009})}\BibitemShut {NoStop}%
\bibitem [{\citenamefont {Cohen-Tannoudji}\ \emph {et~al.}(1998)\citenamefont
  {Cohen-Tannoudji}, \citenamefont {Dupont-Roc},\ and\ \citenamefont
  {Grynberg}}]{cohen-tannoudji_atom-photon_1998}%
  \BibitemOpen
  \bibfield  {author} {\bibinfo {author} {\bibfnamefont {C.}~\bibnamefont
  {Cohen-Tannoudji}}, \bibinfo {author} {\bibfnamefont {J.}~\bibnamefont
  {Dupont-Roc}}, \ and\ \bibinfo {author} {\bibfnamefont {G.}~\bibnamefont
  {Grynberg}},\ }\href@noop {} {{\selectlanguage {en}\emph {\bibinfo {title}
  {Atom-{Photon} {Interactions}: {Basic} {Processes} and {Applications}}}}}\
  (\bibinfo  {publisher} {Wiley},\ \bibinfo {year} {1998})\BibitemShut
  {NoStop}%
\bibitem [{\citenamefont {Berkolaiko}\ and\ \citenamefont
  {Kuipers}(2010)}]{berkolaiko_moments_2010}%
  \BibitemOpen
  \bibfield  {author} {\bibinfo {author} {\bibfnamefont {G.}~\bibnamefont
  {Berkolaiko}}\ and\ \bibinfo {author} {\bibfnamefont {J.}~\bibnamefont
  {Kuipers}},\ }\href {\doibase 10.1088/1751-8113/43/3/035101} {\bibfield
  {journal} {\bibinfo  {journal} {J. Phys. A: Math. Theor.}\ }\textbf {\bibinfo
  {volume} {43}},\ \bibinfo {pages} {035101} (\bibinfo {year}
  {2010})}\BibitemShut {NoStop}%
\bibitem [{\citenamefont {Kuipers}\ and\ \citenamefont
  {Sieber}(2008)}]{kuipers_semiclassical_2008}%
  \BibitemOpen
  \bibfield  {author} {\bibinfo {author} {\bibfnamefont {J.}~\bibnamefont
  {Kuipers}}\ and\ \bibinfo {author} {\bibfnamefont {M.}~\bibnamefont
  {Sieber}},\ }\href {\doibase 10.1103/PhysRevE.77.046219} {\bibfield
  {journal} {\bibinfo  {journal} {Phys. Rev. E}\ }\textbf {\bibinfo {volume}
  {77}},\ \bibinfo {pages} {046219} (\bibinfo {year} {2008})}\BibitemShut
  {NoStop}%
\bibitem [{\citenamefont {Smith}(1960)}]{smith_lifetime_1960}%
  \BibitemOpen
  \bibfield  {author} {\bibinfo {author} {\bibfnamefont {F.~T.}\ \bibnamefont
  {Smith}},\ }\href {\doibase 10.1103/PhysRev.119.2098.4} {\bibfield  {journal}
  {\bibinfo  {journal} {Phys. Rev.}\ }\textbf {\bibinfo {volume} {119}},\
  \bibinfo {pages} {2098} (\bibinfo {year} {1960})}\BibitemShut {NoStop}%
\bibitem [{\citenamefont {Rotter}(2013)}]{Rotter2013}%
  \BibitemOpen
  \bibfield  {author} {\bibinfo {author} {\bibfnamefont {I.}~\bibnamefont
  {Rotter}},\ }\href {\doibase 10.1002/prop.201200054} {\bibfield  {journal}
  {\bibinfo  {journal} {Fortschritte der Physik}\ }\textbf {\bibinfo {volume}
  {61}},\ \bibinfo {pages} {178} (\bibinfo {year} {2013})}\BibitemShut
  {NoStop}%
\bibitem [{\citenamefont {Lyuboshitz}(1977)}]{lyuboshitz_collision_1977}%
  \BibitemOpen
  \bibfield  {author} {\bibinfo {author} {\bibfnamefont {V.~L.}\ \bibnamefont
  {Lyuboshitz}},\ }\href {\doibase 10.1016/0370-2693(77)90058-2} {\bibfield
  {journal} {\bibinfo  {journal} {Physics Letters B}\ }\textbf {\bibinfo
  {volume} {72}},\ \bibinfo {pages} {41} (\bibinfo {year} {1977})}\BibitemShut
  {NoStop}%
\bibitem [{\citenamefont {Zech}\ \emph {et~al.}(2014)\citenamefont {Zech},
  \citenamefont {Mulet}, \citenamefont {Wellens},\ and\ \citenamefont
  {Buchleitner}}]{zech_centrosymmetry_2014}%
  \BibitemOpen
  \bibfield  {author} {\bibinfo {author} {\bibfnamefont {T.}~\bibnamefont
  {Zech}}, \bibinfo {author} {\bibfnamefont {R.}~\bibnamefont {Mulet}},
  \bibinfo {author} {\bibfnamefont {T.}~\bibnamefont {Wellens}}, \ and\
  \bibinfo {author} {\bibfnamefont {A.}~\bibnamefont {Buchleitner}},\ }\href
  {\doibase 10.1088/1367-2630/16/5/055002} {\bibfield  {journal} {\bibinfo
  {journal} {New J. Phys.}\ }\textbf {\bibinfo {volume} {16}},\ \bibinfo
  {pages} {055002} (\bibinfo {year} {2014})}\BibitemShut {NoStop}%
\bibitem [{\citenamefont {Walschaers}\ \emph {et~al.}(2013)\citenamefont
  {Walschaers}, \citenamefont {Diaz}, \citenamefont {Mulet},\ and\
  \citenamefont {Buchleitner}}]{walschaers_optimally_2013}%
  \BibitemOpen
  \bibfield  {author} {\bibinfo {author} {\bibfnamefont {M.}~\bibnamefont
  {Walschaers}}, \bibinfo {author} {\bibfnamefont {J.~F.-d.-C.}\ \bibnamefont
  {Diaz}}, \bibinfo {author} {\bibfnamefont {R.}~\bibnamefont {Mulet}}, \ and\
  \bibinfo {author} {\bibfnamefont {A.}~\bibnamefont {Buchleitner}},\ }\href
  {\doibase 10.1103/PhysRevLett.111.180601} {\bibfield  {journal} {\bibinfo
  {journal} {Phys. Rev. Lett.}\ }\textbf {\bibinfo {volume} {111}},\ \bibinfo
  {pages} {180601} (\bibinfo {year} {2013})}\BibitemShut {NoStop}%
\bibitem [{\citenamefont {Walschaers}\ \emph {et~al.}(2015)\citenamefont
  {Walschaers}, \citenamefont {Mulet}, \citenamefont {Wellens},\ and\
  \citenamefont {Buchleitner}}]{walschaers_statistical_2015}%
  \BibitemOpen
  \bibfield  {author} {\bibinfo {author} {\bibfnamefont {M.}~\bibnamefont
  {Walschaers}}, \bibinfo {author} {\bibfnamefont {R.}~\bibnamefont {Mulet}},
  \bibinfo {author} {\bibfnamefont {T.}~\bibnamefont {Wellens}}, \ and\
  \bibinfo {author} {\bibfnamefont {A.}~\bibnamefont {Buchleitner}},\ }\href
  {\doibase 10.1103/PhysRevE.91.042137} {\bibfield  {journal} {\bibinfo
  {journal} {Phys. Rev. E}\ }\textbf {\bibinfo {volume} {91}},\ \bibinfo
  {pages} {042137} (\bibinfo {year} {2015})}\BibitemShut {NoStop}%
\bibitem [{\citenamefont {Walschaers}\ \emph {et~al.}(2017)\citenamefont
  {Walschaers}, \citenamefont {Buchleitner},\ and\ \citenamefont
  {Fannes}}]{walschaers_currents}%
  \BibitemOpen
  \bibfield  {author} {\bibinfo {author} {\bibfnamefont {M.}~\bibnamefont
  {Walschaers}}, \bibinfo {author} {\bibfnamefont {A.}~\bibnamefont
  {Buchleitner}}, \ and\ \bibinfo {author} {\bibfnamefont {M.}~\bibnamefont
  {Fannes}},\ }\href {http://stacks.iop.org/1367-2630/19/i=2/a=023025}
  {\bibfield  {journal} {\bibinfo  {journal} {New Journal of Physics}\ }\textbf
  {\bibinfo {volume} {19}},\ \bibinfo {pages} {023025} (\bibinfo {year}
  {2017})}\BibitemShut {NoStop}%
\bibitem [{\citenamefont {Ortega}\ \emph {et~al.}(2015)\citenamefont {Ortega},
  \citenamefont {Vyas},\ and\ \citenamefont {Benet}}]{ortega_quantum_2015}%
  \BibitemOpen
  \bibfield  {author} {\bibinfo {author} {\bibfnamefont {A.}~\bibnamefont
  {Ortega}}, \bibinfo {author} {\bibfnamefont {M.}~\bibnamefont {Vyas}}, \ and\
  \bibinfo {author} {\bibfnamefont {L.}~\bibnamefont {Benet}},\ }\href
  {\doibase 10.1002/andp.201500140} {\bibfield  {journal} {\bibinfo  {journal}
  {Ann. Phys.}\ }\textbf {\bibinfo {volume} {527}},\ \bibinfo {pages} {748}
  (\bibinfo {year} {2015})}\BibitemShut {NoStop}%
\bibitem [{\citenamefont {Ortega}\ \emph {et~al.}(2016)\citenamefont {Ortega},
  \citenamefont {Stegmann},\ and\ \citenamefont {Benet}}]{PhysRevE.94.042102}%
  \BibitemOpen
  \bibfield  {author} {\bibinfo {author} {\bibfnamefont {A.}~\bibnamefont
  {Ortega}}, \bibinfo {author} {\bibfnamefont {T.}~\bibnamefont {Stegmann}}, \
  and\ \bibinfo {author} {\bibfnamefont {L.}~\bibnamefont {Benet}},\ }\href
  {\doibase 10.1103/PhysRevE.94.042102} {\bibfield  {journal} {\bibinfo
  {journal} {Phys. Rev. E}\ }\textbf {\bibinfo {volume} {94}},\ \bibinfo
  {pages} {042102} (\bibinfo {year} {2016})}\BibitemShut {NoStop}%
\bibitem [{\citenamefont {Walschaers}(2016)}]{Walschaers_Thesis}%
  \BibitemOpen
  \bibfield  {author} {\bibinfo {author} {\bibfnamefont {M.}~\bibnamefont
  {Walschaers}},\ }\emph {\bibinfo {title} {Efficient Quantum Transport}},\
  \href@noop {} {Ph.D. thesis},\ \bibinfo  {school} {Albert-Ludwigs
  Universit{\"a}t Freiburg and KU Leuven} (\bibinfo {year} {2016})\BibitemShut
  {NoStop}%
\bibitem [{Note2()}]{Note2}%
  \BibitemOpen
  \bibinfo {note} {Also note the negative value of $\tau _{\protect \rm in
  \rightarrow out}$ at $E\approx 1.5$, which hints at the fact that (\ref
  {eq:DefTauDenkIK}) actually quantifies a phase shift rather than a genuine
  time.}\BibitemShut {Stop}%
\bibitem [{Note3()}]{Note3}%
  \BibitemOpen
  \bibinfo {note} {We will henceforth omit the index ``${\protect \rm
  in}\rightarrow {\protect \rm out}$'', for ease of notation.}\BibitemShut
  {Stop}%
\bibitem [{\citenamefont {L{\'o}pez}\ \emph {et~al.}(1981)\citenamefont
  {L{\'o}pez}, \citenamefont {Mello},\ and\ \citenamefont
  {Seligman}}]{Lopez1981}%
  \BibitemOpen
  \bibfield  {author} {\bibinfo {author} {\bibfnamefont {G.}~\bibnamefont
  {L{\'o}pez}}, \bibinfo {author} {\bibfnamefont {P.~A.}\ \bibnamefont
  {Mello}}, \ and\ \bibinfo {author} {\bibfnamefont {T.~H.}\ \bibnamefont
  {Seligman}},\ }\href {\doibase 10.1007/BF01414267} {\bibfield  {journal}
  {\bibinfo  {journal} {Zeitschrift f{\"u}r Physik A Atoms and Nuclei}\
  }\textbf {\bibinfo {volume} {302}},\ \bibinfo {pages} {351} (\bibinfo {year}
  {1981})}\BibitemShut {NoStop}%
\bibitem [{\citenamefont {Tomsovic}\ and\ \citenamefont
  {Ullmo}(1994)}]{tomsovic_chaos-assisted_1994}%
  \BibitemOpen
  \bibfield  {author} {\bibinfo {author} {\bibfnamefont {S.}~\bibnamefont
  {Tomsovic}}\ and\ \bibinfo {author} {\bibfnamefont {D.}~\bibnamefont
  {Ullmo}},\ }\href {\doibase 10.1103/PhysRevE.50.145} {\bibfield  {journal}
  {\bibinfo  {journal} {Phys. Rev. E}\ }\textbf {\bibinfo {volume} {50}},\
  \bibinfo {pages} {145} (\bibinfo {year} {1994})}\BibitemShut {NoStop}%
\bibitem [{\citenamefont {Leyvraz}\ and\ \citenamefont
  {Ullmo}(1996)}]{leyvraz_level_1996}%
  \BibitemOpen
  \bibfield  {author} {\bibinfo {author} {\bibfnamefont {F.}~\bibnamefont
  {Leyvraz}}\ and\ \bibinfo {author} {\bibfnamefont {D.}~\bibnamefont
  {Ullmo}},\ }\href {\doibase 10.1088/0305-4470/29/10/030} {\bibfield
  {journal} {\bibinfo  {journal} {J. Phys. A: Math. Gen.}\ }\textbf {\bibinfo
  {volume} {29}},\ \bibinfo {pages} {2529} (\bibinfo {year}
  {1996})}\BibitemShut {NoStop}%
\bibitem [{Note4()}]{Note4}%
  \BibitemOpen
  \bibinfo {note} {More details on this model can be found in Chapter 4 of
  \cite {Walschaers_Thesis}.}\BibitemShut {Stop}%
\bibitem [{\citenamefont {Erg{\"u}n}\ and\ \citenamefont
  {Fyodorov}(2003)}]{ergun_level_2003}%
  \BibitemOpen
  \bibfield  {author} {\bibinfo {author} {\bibfnamefont {G.}~\bibnamefont
  {Erg{\"u}n}}\ and\ \bibinfo {author} {\bibfnamefont {Y.~V.}\ \bibnamefont
  {Fyodorov}},\ }\href {\doibase 10.1103/PhysRevE.68.046124} {\bibfield
  {journal} {\bibinfo  {journal} {Phys. Rev. E}\ }\textbf {\bibinfo {volume}
  {68}},\ \bibinfo {pages} {046124} (\bibinfo {year} {2003})}\BibitemShut
  {NoStop}%
\bibitem [{\citenamefont {Brouwer}(1997)}]{brouwer_random-matrix_1997}%
  \BibitemOpen
  \bibfield  {author} {\bibinfo {author} {\bibfnamefont {P.~W.}\ \bibnamefont
  {Brouwer}},\ }\emph {\bibinfo {title} {On the {Random}-{Matrix} {Theory} of
  {Quantum} {Transport}}},\ \href
  {http://www.lorentz.leidenuniv.nl/beenakker/theses/brouwer/brouwer.html}
  {\bibinfo {type} {{PhD} {Thesis}}},\ \bibinfo  {school} {Leiden Univerisity},
  \bibinfo {address} {Leiden} (\bibinfo {year} {1997})\BibitemShut {NoStop}%
\bibitem [{\citenamefont {Bohigas}\ \emph {et~al.}(1993)\citenamefont
  {Bohigas}, \citenamefont {Tomsovic},\ and\ \citenamefont
  {Ullmo}}]{bohigas_manifestations_1993}%
  \BibitemOpen
  \bibfield  {author} {\bibinfo {author} {\bibfnamefont {O.}~\bibnamefont
  {Bohigas}}, \bibinfo {author} {\bibfnamefont {S.}~\bibnamefont {Tomsovic}}, \
  and\ \bibinfo {author} {\bibfnamefont {D.}~\bibnamefont {Ullmo}},\ }\href
  {\doibase 10.1016/0370-1573(93)90109-Q} {\bibfield  {journal} {\bibinfo
  {journal} {Phys. Rep.}\ }\textbf {\bibinfo {volume} {223}},\ \bibinfo {pages}
  {43} (\bibinfo {year} {1993})}\BibitemShut {NoStop}%
\bibitem [{\citenamefont {Scholak}\ \emph {et~al.}(2011)\citenamefont
  {Scholak}, \citenamefont {de~Melo}, \citenamefont {Wellens}, \citenamefont
  {Mintert},\ and\ \citenamefont {Buchleitner}}]{scholak_efficient_2011}%
  \BibitemOpen
  \bibfield  {author} {\bibinfo {author} {\bibfnamefont {T.}~\bibnamefont
  {Scholak}}, \bibinfo {author} {\bibfnamefont {F.}~\bibnamefont {de~Melo}},
  \bibinfo {author} {\bibfnamefont {T.}~\bibnamefont {Wellens}}, \bibinfo
  {author} {\bibfnamefont {F.}~\bibnamefont {Mintert}}, \ and\ \bibinfo
  {author} {\bibfnamefont {A.}~\bibnamefont {Buchleitner}},\ }\href {\doibase
  10.1103/PhysRevE.83.021912} {\bibfield  {journal} {\bibinfo  {journal} {Phys.
  Rev. E}\ }\textbf {\bibinfo {volume} {83}},\ \bibinfo {pages} {021912}
  (\bibinfo {year} {2011})}\BibitemShut {NoStop}%
\bibitem [{\citenamefont {Engel}\ \emph {et~al.}(2007)\citenamefont {Engel},
  \citenamefont {Calhoun}, \citenamefont {Read}, \citenamefont {Ahn},
  \citenamefont {Man{\v c}al}, \citenamefont {Cheng}, \citenamefont
  {Blankenship},\ and\ \citenamefont {Fleming}}]{engel_evidence_2007}%
  \BibitemOpen
  \bibfield  {author} {\bibinfo {author} {\bibfnamefont {G.~S.}\ \bibnamefont
  {Engel}}, \bibinfo {author} {\bibfnamefont {T.~R.}\ \bibnamefont {Calhoun}},
  \bibinfo {author} {\bibfnamefont {E.~L.}\ \bibnamefont {Read}}, \bibinfo
  {author} {\bibfnamefont {T.-K.}\ \bibnamefont {Ahn}}, \bibinfo {author}
  {\bibfnamefont {T.}~\bibnamefont {Man{\v c}al}}, \bibinfo {author}
  {\bibfnamefont {Y.-C.}\ \bibnamefont {Cheng}}, \bibinfo {author}
  {\bibfnamefont {R.~E.}\ \bibnamefont {Blankenship}}, \ and\ \bibinfo {author}
  {\bibfnamefont {G.~R.}\ \bibnamefont {Fleming}},\ }\href {\doibase
  10.1038/nature05678} {\bibfield  {journal} {\bibinfo  {journal} {Nature}\
  }\textbf {\bibinfo {volume} {446}},\ \bibinfo {pages} {782} (\bibinfo {year}
  {2007})}\BibitemShut {NoStop}%
\bibitem [{\citenamefont {Hildner}\ \emph {et~al.}(2013)\citenamefont
  {Hildner}, \citenamefont {Brinks}, \citenamefont {Nieder}, \citenamefont
  {Cogdell},\ and\ \citenamefont {Hulst}}]{hildner_quantum_2013}%
  \BibitemOpen
  \bibfield  {author} {\bibinfo {author} {\bibfnamefont {R.}~\bibnamefont
  {Hildner}}, \bibinfo {author} {\bibfnamefont {D.}~\bibnamefont {Brinks}},
  \bibinfo {author} {\bibfnamefont {J.~B.}\ \bibnamefont {Nieder}}, \bibinfo
  {author} {\bibfnamefont {R.~J.}\ \bibnamefont {Cogdell}}, \ and\ \bibinfo
  {author} {\bibfnamefont {N.~F.~v.}\ \bibnamefont {Hulst}},\ }\href {\doibase
  10.1126/science.1235820} {\bibfield  {journal} {\bibinfo  {journal}
  {Science}\ }\textbf {\bibinfo {volume} {340}},\ \bibinfo {pages} {1448}
  (\bibinfo {year} {2013})}\BibitemShut {NoStop}%
\end{thebibliography}%

\end{document}